\documentclass[twocolumn,trackchanges]{aastex701}
\usepackage{amsmath}
\hypersetup{linkcolor=red,citecolor=blue,filecolor=cyan,urlcolor=magenta}

\graphicspath{{./}{figures/}}

\begin{document}

\title{Limiting Eccentricity in Restricted Hierarchical Three-Body Systems with Short-Range Forces}

\correspondingauthor{Xiumin Huang}
\email{xm\_huang@sjtu.edu.cn}
\correspondingauthor{Dong Lai}
\email{donglai@sjtu.edu.cn}

\author{Xiumin Huang}
\affiliation{Tsung-Dao Lee Institute, Shanghai Jiao-Tong University, Shanghai, 201210, China}
\email{xm\_huang@sjtu.edu.cn}  

\author{Dong Lai}
\affiliation{Tsung-Dao Lee Institute, Shanghai Jiao-Tong University, Shanghai, 201210, China}
\affiliation{Center for Astrophysics and Planetary Science, Department of Astronomy, Cornell University, Ithaca, NY 14853, USA}
\email{donglai@sjtu.edu.cn}

\author{Bin Liu}
\affiliation{Institute for Astronomy, School of Physics, Zhejiang University, 310058 Hangzhou, China}
\email{liubin23@zju.edu.cn}

\begin{abstract}
A hierarchical three-body model can be widely applied to diverse astrophysical settings, from satellite-planet-star systems to binaries around supermassive black holes. The octupole-order perturbation on the inner binary from the tertiary can induce extreme eccentricities and cause orbital flips of the binary, but short-range forces such as those due to General Relativity (GR) may suppress extreme eccentricity excitations. In this paper, we consider restricted hierarchical three-body systems, where the inner binary has a test-mass component. We investigate the maximum possible eccentricity (called ``limiting eccentricity") attainable by the inner binary under the influence of the tertiary perturbations and GR effect. In systems with sufficiently high hierarchy, the double averaging (DA) model is a good approximation; we show that the orbits which can flip under the octupole-order perturbation reach the same limiting eccentricity, which can be calculated analytically using the quadrupole-order Hamiltonian. In systems with moderate hierarchy, DA breaks down and the so-called Brown Hamiltonian is often introduced as a correction term; we show that this does not change the limiting eccentricity. Finally, we employ the single averaging (SA) model and find that the limiting eccentricity in the SA model is higher than the one in the DA model. We derive an analytical scaling for the modified limiting eccentricity in the SA model.
\end{abstract}

\keywords{gravitation -- celestial mechanics }

\section{Introduction}
Hierarchical three-body systems, where the semi-major axis of the inner binary, $a$, is much less than that of the outer tertiary perturber, $a_p$, exist in various astrophysical settings, such as planetary satellites perturbed by the Sun, a planet around its host star perturbed by a distant companion, and stellar binaries in a nuclear star cluster perturbed by the supermassive black hole. \citet{von1910application} and \citet{lidov1962evolution} (see also \citealp{kozai1962secular}) found that in a restricted hierarchical system (where the inner binary includes a test-mass component), when the initial inclination of the inner circular orbit relative to the outer orbit is between $40^{\circ}$ and $140^{\circ}$, the leading quadrupole-order perturbation from the tertiary can induce coupled eccentricity and inclination oscillations of the inner binary. This phenomenon is referred to as the von Zeipel-Lidov-Kozai (ZLK) effect \citep{ito2019lidov}. The large eccentricity excitations associated with large initial inclinations ($i \sim 90^{\circ}$) have found many applications in recent years (e.g. \citealp{naoz2016eccentric}). For examples, the high eccentricity obtained from this effect can lead to the formation of hot Jupiters via high-eccentricity migration (e.g. \citealp{wu2003planet,fabrycky2007shrinking,petrovich2015steady,anderson2016formation,dawson2018origins,vick2019chaotic}), facilitate tertiary-induced binary black hole mergers (e.g. \citealp{antonini2012secular,liu2017spin,liu2018black}) and tidal disruption events (e.g. \citealp{melchor2023tidal}).

In three-body systems with sufficiently high hierarchy, the inner and outer orbits can both be averaged in the Hamiltonian (e.g., by von Zeipel method) to study the secular effect; this is called double averaging (DA) approximation. Considering the binary-tertiary Hamiltonian only up to the quadrupole order in the restricted three-body problem, the $z$-component (i.e., along the direction of the outer orbital angular momentum) of the inner orbital angular momentum is conserved and the system has one degree of freedom. Therefore, the maximum eccentricity of the inner orbit can be derived analytically. For an initial circular orbit, the maximum eccentricity $e_{\rm max}$ is a function of initial inclination $i_0$, 
\begin{equation}
    {e_{\max }} = \sqrt {1 - \frac{5}{3}{{\cos }^2}{i_{0}}},
\end{equation}
which gives $e_{\max}=1$ when $i_0=90^{\circ}$. 

Including the octupole-order binary-tertiary interaction in the Hamiltonian \citep{ford2000secular} brings richer dynamics, and may lead to chaos due to the introduction of a new degree of freedom. The strength of the octupole-order effect is measured by the dimensionless parameter
\begin{equation}\label{eq_epsilonoct}
    \epsilon_{\rm Oct}=\frac{a}{a_p}\frac{e_p}{1-e_p^2},
\end{equation}
where $e_p$ is the eccentricity of the outer orbit. When $\epsilon_{\rm Oct}$ is not negligible (e.g., $\epsilon_{\rm Oct} \gtrsim 0.01$), the eccentricity of the inner binary can be excited to extremely close to unity and the orbit can flip between prograde and retrograde even when the initial inclination is not too close to $90^{\circ}$ \citep{naoz2011hot,katz2011long,lithwick2011eccentric,munoz2016formation,lei2022systematic,klein2024hierarchical}.

However, if the eccentricity is excited to a large value, the test particle can be very close to the central body at the pericenter, and various short-range forces (SRFs), such as those due to General Relativity (GR), tidal and rotational distortions, can become significant, thereby suppressing eccentricity growth \citep{holman1997chaotic,wu2003planet,fabrycky2007shrinking}. This suppression arises because of the apsidal precessions (associated with the SRFs) de-tune the secular resonance for eccentricity
driving when the precession timescale becomes comparable to the ZLK timescale. The ZLK timescale is given by  \citep{antognini2015timescales,liu2018black}
\begin{equation}
    t_{\rm{ZKL}} = \frac{P_{\rm in}}{2\pi} \frac{m_*}{m_p}  \left(\frac{a_p\sqrt{1-e_p^2}}{a} \right)^{3},
\end{equation}
where $m_*$ and $m_p$ are the masses of the central object and the tertiary, respectively, and $P_{\rm in}$ is the inner orbital period. In the quadrupole approximation of the three-body Hamiltonian, the effect of SRFs can be analytically solved to determine the maximum eccentricity as a function of the initial inclination $i_0$ (\citealp{fabrycky2007shrinking}; see also \citealp{anderson2017eccentricity} for the general
non-test-mass case). \citet{liu2015suppression} showed that even in systems where the octupole terms are significant, the SRFs can suppress eccentricity growth. In particular, they found from numerical examples that in the presence of SRFs, there exists a limiting eccentricity $e_{\rm lim}$ (equal to the maximum eccentricity at $i_0=90^\circ$) that the inner binary can attain,
regardless of the octupole strength; above a critical inclination angle (whose value depends on $\epsilon_{\rm Oct}$; see \citealp{munoz2016formation}),
the maximum possible eccentricity appears to be equal to $e_{\rm lim}$. In this paper, we revisit this problem analytically, and quantify the empirical finding of \citet{liu2015suppression}. For concreteness, we include only the GR effect as an example of the SRFs, but our analysis also applies to other types of SRFs.

A necessary condition for the validity of the DA model is $P_{\rm in} \ll P_{\rm out} \ll t_{\rm ZLK}$, where $P_{\rm out}$ is the outer orbital period. In addition, since the timescale of eccentricity variation near $e_{\max}$ is $t_{\rm ZLK}\sqrt{1-e_{\max}^2}$ (e.g. \citealp{liu2018black}), the validity of the DA model also requires 
\begin{equation}\label{eq_davalid}
    t_{\rm ZLK}\sqrt{1-e_{\max}^2} \gtrsim P_{\rm out}.
\end{equation}
When $P_{\rm out}$ becomes larger than $t_{\rm ZLK}\sqrt{1-e_{\max}^2}$, the variation of the inner orbital angular momentum within one outer orbital period is not negligible, and the DA approximation breaks down \citep{liu2018black,grishin2018quasi,lei2025extensions}. To address this problem, the Brown Hamiltonian (BH) can be introduced as a correction term to the DA Hamiltonian \citep{luo2016double, tremaine2023hamiltonian}, or one can employ the single averaging (SA) approximation, where the Hamiltonian is only averaged over the inner orbital period. The SA model is valid when the timescale for eccentricity evolution around $e_{\max}$ is longer than the inner orbital period, i.e.
\begin{equation}
    t_{\rm ZLK}\sqrt{1-e_{\max}^2} \gtrsim P_{\rm in}.
\end{equation}
\citet{liu2018black} found examples where the maximum eccentricity in the SA model is higher than the one in the DA model. But whether there exists a limiting eccentricity in the SA model or in the ``DA + BH" model is unclear.

In this work, we investigate the maximum eccentricity under the influence of SRFs (using GR as an example of SRFs) for different models of 3-body hierarchical restricted systems. In Section \ref{s2} we study the maximum eccentricity in the DA model, which is a good approximation for systems with sufficiently high hierarchy. Then we study the maximum eccentricity in moderate hierarchical systems. We employ the DA model with the BH correction and the SA model in Section \ref{s3} and Section \ref{s4}, respectively. We summarize our findings in Section \ref{s5}.

\section{Double averaging (DA) model}\label{s2}
In a hierarchical restricted three-body system, a particle orbits a central object of mass $m_*$, with a distant perturber of mass $m_p$. Let $a$ and $a_p$ be the semi-major axes of the inner and outer orbits, respectively. The secular Hamiltonian of the system (double averaged over the inner and outer orbits), including the gravitational interaction between the inner binary and the tertiary up to the octupole order and the post-Newtonian potential associated with periastron advance of the inner orbit \citep{krymolowski1999studies,ford2000secular,naoz2013secular,eggleton2001orbital,liu2015suppression}, can be written as
\begin{equation}\label{equa0}
\begin{aligned}
    {\cal H} &= \Phi_0({\hat{\cal H}_{\rm Quad}} + {\hat{\cal H}_{\rm Oct}} + \hat{\Phi}_{\rm GR}),
\end{aligned}
\end{equation}
where
\begin{equation}
\begin{aligned}
    {\hat{\cal H}_{\rm Quad}}{\rm{ = }} &- \frac{1}{8}\{ 2 + 3{e^2} - [3 + 12{e^2} \\
    &- \frac{{15}}{2}{e^2}(1 + \cos 2\omega )](1 - {\theta ^2})\}
\end{aligned}
\end{equation}
is the dimensionless quadrupole-order term,
\begin{equation}
\begin{aligned}
    {\hat{\cal H}_{\rm Oct}} =  &- \frac{{15}}{{512}}{\epsilon _{{\rm{Oct}}}}e\{ (4 + 3{e^2})[(1 - 11\theta  - 5{\theta ^2}\\
 &+ 15{\theta ^3})\cos (\omega  - \Omega ) + (1 + 11\theta  - 5{\theta ^2}\\
 &- 15{\theta ^3})\cos (\omega  + \Omega )] - 35{e^2}(1 - {\theta ^2})[(1\\
 &- \theta )\cos (3\omega  - \Omega ) + (1 + \theta )\cos (3\omega  + \Omega )]\}
\end{aligned}   
\end{equation}
is the octupole-order term, and
\begin{equation}
    \hat{\Phi}_{\rm GR}  =  - \frac{{{\epsilon _{\rm GR}}}}{{\sqrt {1 - {e^2}} }}
\end{equation}
is the potential due to GR. In the above equations,
\begin{equation}
    {\Phi _0} = \frac{{{G}{m_p}{{a^{\rm{2}}}}}}{{a_p^3{{(1 - e_p^2)}^{3/2}}}}
\end{equation}
measures the strength of the quadrupole potential ($G$ is the gravitational constant), $e$, $\omega$ and $\Omega$ represent the inner orbital eccentricity, argument of periapse and longitude of ascending node, respectively, and $\theta \equiv \cos i$, with $i$ the inclination angle between the inner and outer orbits. The dimensionless coefficient ${\epsilon _{{\rm{Oct}}}}$ (measuring the importance of the octupole term compared to the quadrupole-order term) is given by Eq. (\ref{eq_epsilonoct}), and ${\epsilon _{{\rm{GR}}}}$ (measuring the importance of the GR term) is given by
\begin{equation}
   {\epsilon _{\rm GR}} = \frac{{3{G}{m_*}^2a_p^3{{(1 - e_p^2)}^{3/2}}}}{{{a^4}{c^2}{m_p}}},
\end{equation}
where $c$ is the speed of light. The total energy of the system $\cal H$ is conserved, and the semi-major axes $a$ and $a_p$ are constants because the orbital phases of inner and outer orbits have been removed in the secular Hamiltonian.

To derive the equations of motion, we introduce the following canonical variables:
\begin{equation}\label{eq_H}
\begin{aligned}
{J}&=\sqrt{1-{{e}^{2}}},\; 
\quad{{j}}={\omega},\\
{{H}}&={{J}}\cos {{i}},\;  \quad{{h}}={{\Omega }},
\end{aligned}
\end{equation}
where $J$ is the normalized angular momentum of the inner orbit, and $H$ is the $z$-component (i.e., along the outer binary angular momentum axis). The equations of motion are 
\begin{equation}\label{equa_motion}
\begin{aligned}
&\frac{{\rm d}{{j}}}{{\rm d}t}=\frac{\partial \mathcal{H}}{\partial {{J}}},\quad \frac{{\rm d}{{J}}}{{\rm d}t}=-\frac{\partial \mathcal{H}}{\partial {{j}}}, \\
&\frac{{\rm d}{{h}}}{{\rm d}t}=\frac{\partial \mathcal{H}}{\partial {{H}}},\quad \frac{{\rm d}{{H}}}{{\rm d}t}=-\frac{\partial \mathcal{H}}{\partial {{h}}}.
\end{aligned}
\end{equation}

In the following, we study the maximum eccentricity attained by the inner orbit under the quadrupole-order and the octupole-order approximation, respectively.

\subsection{The quadrupole-order result summary}
\begin{figure}
\centering
{\includegraphics[width=0.49\textwidth]{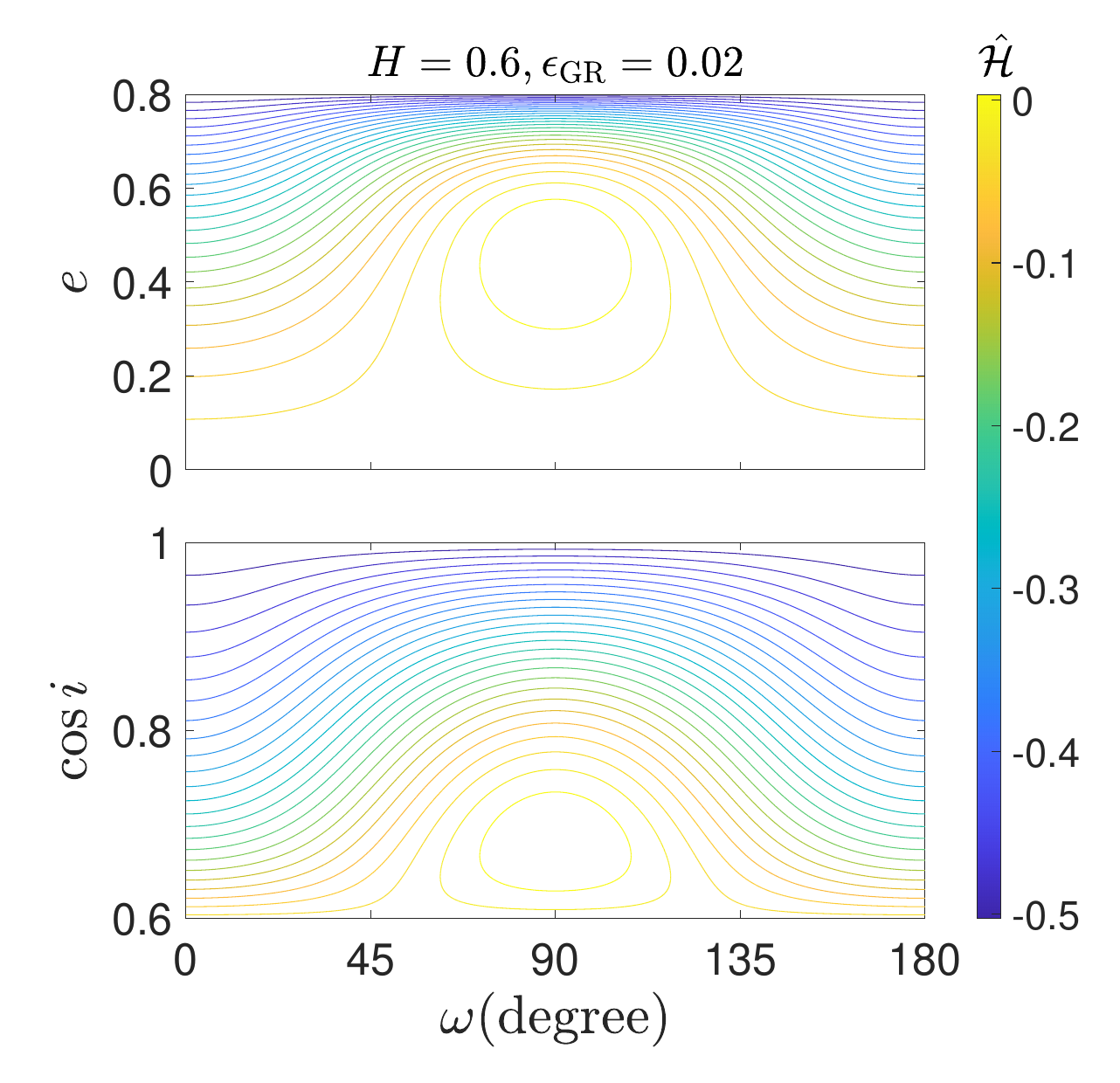}}
\caption{Level curves of (normalized) Hamiltonian for the ``quadrupole + GR" model, with $\epsilon_{\rm GR}=0.02$ and the (normalized) $z$-component of the inner orbital angular momentum $H=0.6$ (see Eq. \ref{eq_H}). The color bars represent different values of Hamiltonian $\hat{\cal{H}}={\cal H}/\Phi_0$.}
\label{fig0}
\end{figure}

In the quadrupole-order approximation ($\epsilon _{\rm {Oct}}=0$), the dynamic is analytical. Considering the inner orbit with the initial values $e_0$, $i_0$ and $\omega_0$, the conservation of $\cal H$ gives
\begin{equation}
\begin{aligned}\label{Hquad}
    &{\hat{\cal H}_{{\rm{Quad}}}}({\omega _0},{e_0},{i_0}) + {\hat{\Phi} _{\rm GR}}({e_0}) \\
    = &{\hat{\cal H}_{{\rm{Quad}}}}({\omega _{\rm emax }},{e_{\max }},{i_{\rm emax }}) + {\hat{\Phi} _{\rm GR}}({e_{\max }}),
\end{aligned}
\end{equation}
and the conservation of the $z$-component of inner orbital angular momentum gives
\begin{equation}
    H=\sqrt {1 - e_0^2} \cos {i_0} = \sqrt {1 - e_{\max }^2} \cos {i_{\rm emax }},
\end{equation}
where $e_{\max}$ is the maximum eccentricity, and the subscript ``$\rm emax$" represents the corresponding orbital elements when $e=e_{\max}$. Note that for a given $e_0$, the maximum eccentricity is always achieved when the initial angle $\omega_0=0$ or $\pi$; at  $e=e_{\max}$, the argument of periapse is always $\omega_{\rm emax}= \pm \pi/2$. The phase portrait for the ``quadrupole + GR'' model in Fig. \ref{fig0} is shown as an example, where the $z$-component of inner orbital angular momentum $H$ is set to 0.6 and $\epsilon_{\rm GR}$ is set to 0.02. 
By applying $\omega_0=0$ and $\omega_{\rm emax}=\pm \pi/2$ in Eq. (\ref{Hquad}), we obtain the analytical expression of the maximum eccentricity:
\begin{equation}\label{equa1}
\begin{aligned}
    & 2{e_{0}}^2 + \frac{{8{\epsilon _{\rm {GR}}}}}{{3\sqrt {1 - {e_{0}}^2} }} - \frac{{8{\epsilon _{\rm {GR}}}}}{{3\sqrt {1 - e_{\max }^2} }} + 3e_{\max }^2 \\
    =& \frac{{5e_{\max }^2}}{{1 - e_{\max }^2}}(1 - {e_{0}}^2){\cos ^2}{i_{0}}.
\end{aligned}
\end{equation}

\begin{figure}
\centering
{\includegraphics[width=0.49\textwidth]{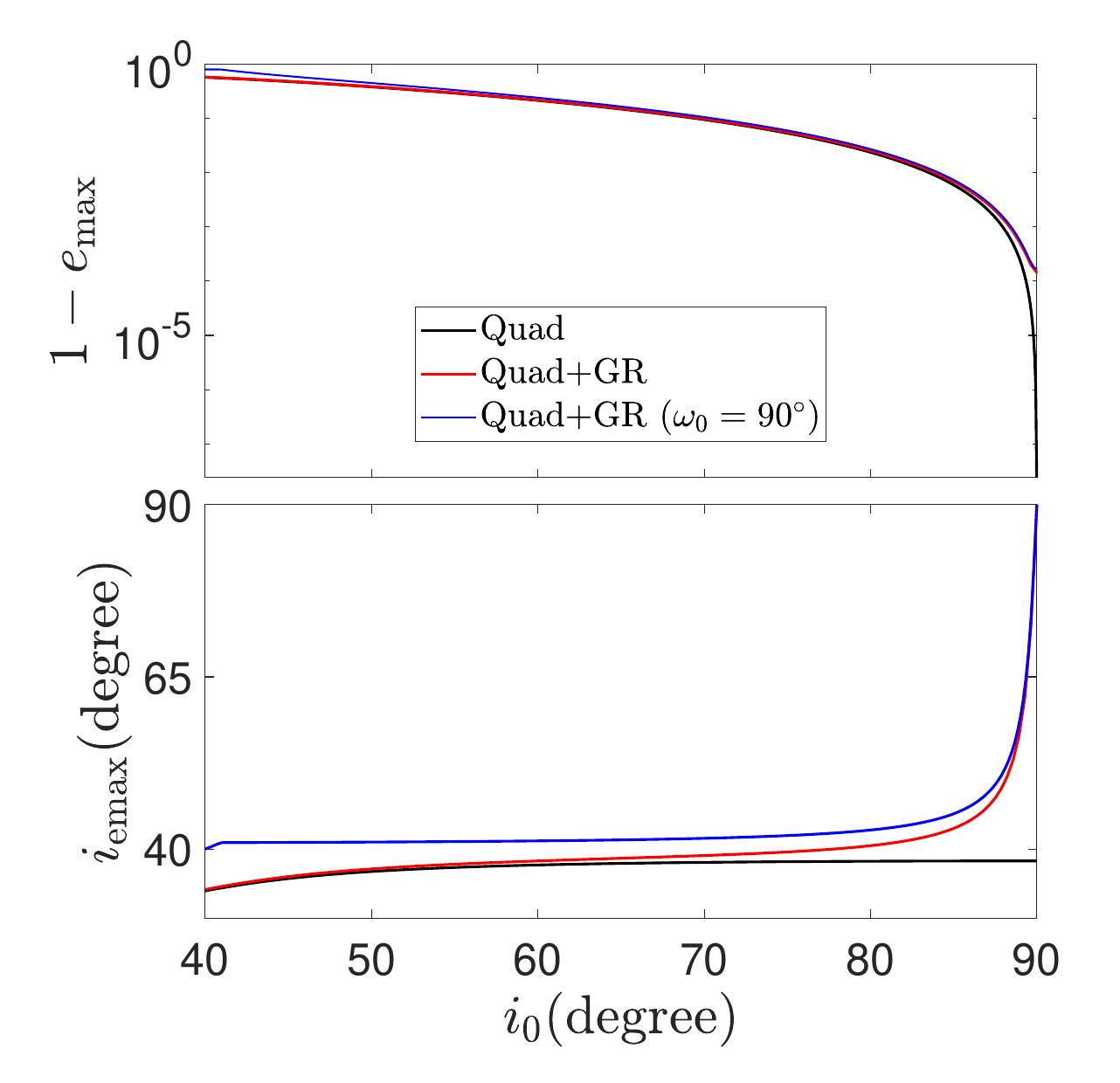}}
\caption{Top panel: The maximum eccentricity that the inner orbit can achieve as a function of the initial inclination $i_{0}$, with the initial eccentricity set to $e_{0}=0.2$. The black line is for the pure quadrupole model without GR. The red line shows the analytical result given by Eq. (\ref{equa1}). The blue line shows the maximum eccentricity with the initial $\omega_0=90^{\circ}$. Both the red and blue lines correspond to the ``quadrupole + GR'' model with $\epsilon_{\rm {GR}}=0.02$. Bottom panel: The corresponding inclination when the inner orbit reaches $e_{\max}$.}
\label{fig1}
\end{figure}

Fig. \ref{fig1} shows the maximum eccentricity that the inner orbit can achieve as a function of the initial inclination $i_{0}$, with the initial eccentricity set to $e_0=0.2$. Without GR (i.e., $\epsilon_{\rm GR}=0$), the maximum eccentricity is denoted by the black line; when $i_0=90^{\circ}$, $e_{\max}=1$. For $\epsilon_{\rm GR}=0.02$, $e_{\max}$ obtained from Eq. (\ref{equa1}) is denoted by the red line, and $e_{\max}$ with initial angle $\omega_0=90^{\circ}$ is denoted by the blue line. As expected, $e_{\max}$ in the blue line is smaller than the one in the red line. The bottom panel of Fig. \ref{fig1} shows the orbital inclination at $e=e_{\max}$. This $i_{\rm emax}$ is actually the minimal inclination that satisfies the conservation of $H$.

According to Fig. \ref{fig1}, as $i_0$ approaches $90^{\circ}$, $e_{\max}$ approaches a limiting value, denoted by $e_{\rm lim}$, i.e., $e_{\rm lim}$ satisfies
\begin{equation}\label{elim}
     2{e_{0}}^2 + \frac{{8{\epsilon _{\rm {GR}}}}}{{3\sqrt {1 - {e_{0}}^2} }} - \frac{{8{\epsilon _{\rm {GR}}}}}{{3\sqrt {1 - e_{\rm lim}^2} }} + 3e_{\rm lim}^2=0.
\end{equation}
For $e_0 \simeq 0$ and $\epsilon_{\rm GR} \ll 1$, this reduces to
\begin{equation}\label{elim_simplify}
\left( 1-e_{\rm lim}^2 \right)^{1/2} \simeq \frac{8}{9} \epsilon_{\rm GR}.
\end{equation}
We will show in Section \ref{sec_oct} that the maximum eccentricity of flipping orbits is $e_{\rm lim}$ in the ``octupole + GR" model.

\subsection{Effect of the octupole term}\label{sec_oct}
It is well recognized that the octupole term of the tertiary perturbation can significantly influence the dynamics of the inner binary. When $\epsilon _{\rm Oct}$ is sufficiently large, the inner orbit can attain extreme eccentricity and undergo orbital flips even for modest initial inclination angles \citep{naoz2011hot}. Nevertheless, \citet{liu2015suppression} showed from numerical examples that the GR effect always sets an upper limit to $e_{\max}$ regardless of the octupole strength $\epsilon _{\rm Oct}$; increasing $\epsilon _{\rm Oct}$ only increases the initial inclination window for extreme eccentricity excitation [see \citet{munoz2016formation} for a fitting formula for the inclination window]. 

\begin{figure*}
\centering
{\includegraphics[width=1\textwidth]{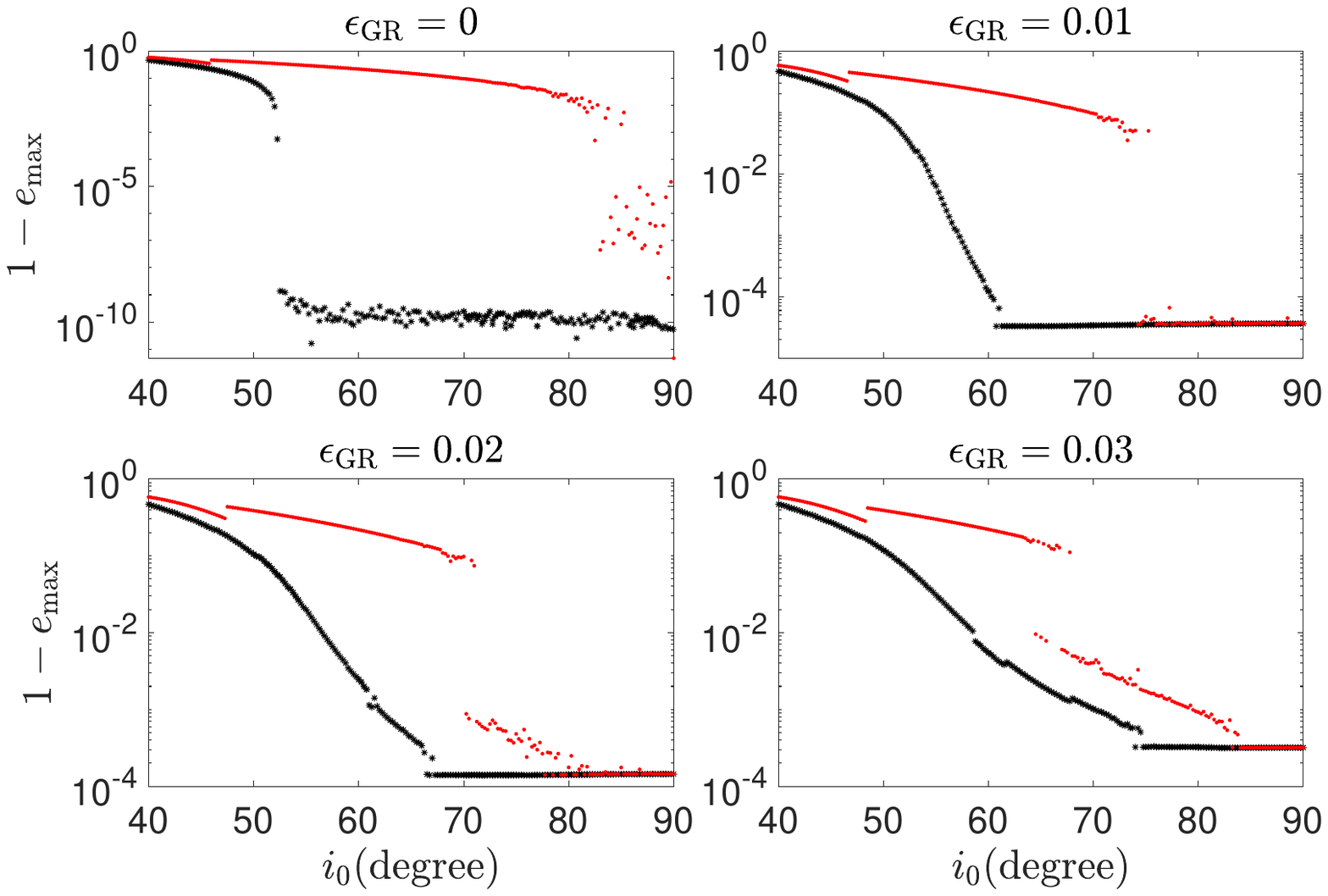}}
\caption{The maximum eccentricity $e_{\max}$ as a function of initial inclination $i_0$ for different values of $\epsilon_{\rm GR}$, with the initial eccentricity set to $e_0=0.2$ and $\epsilon_{\rm Oct}$ set to 0.02. The red dots are the results with the initial $\omega_0=0$ and $\Omega_0=0$. The black asterisks are the results obtained by scanning $\omega_0$ and $\Omega_0$ in the range $[0,2\pi]$. These results are obtained by integrating Eq. (\ref{equa_motion}) for 1500 ZLK cycles for each parameter set.}
\label{figOCT1}
\end{figure*}

\begin{figure*}
\centering
{\includegraphics[width=1\textwidth]{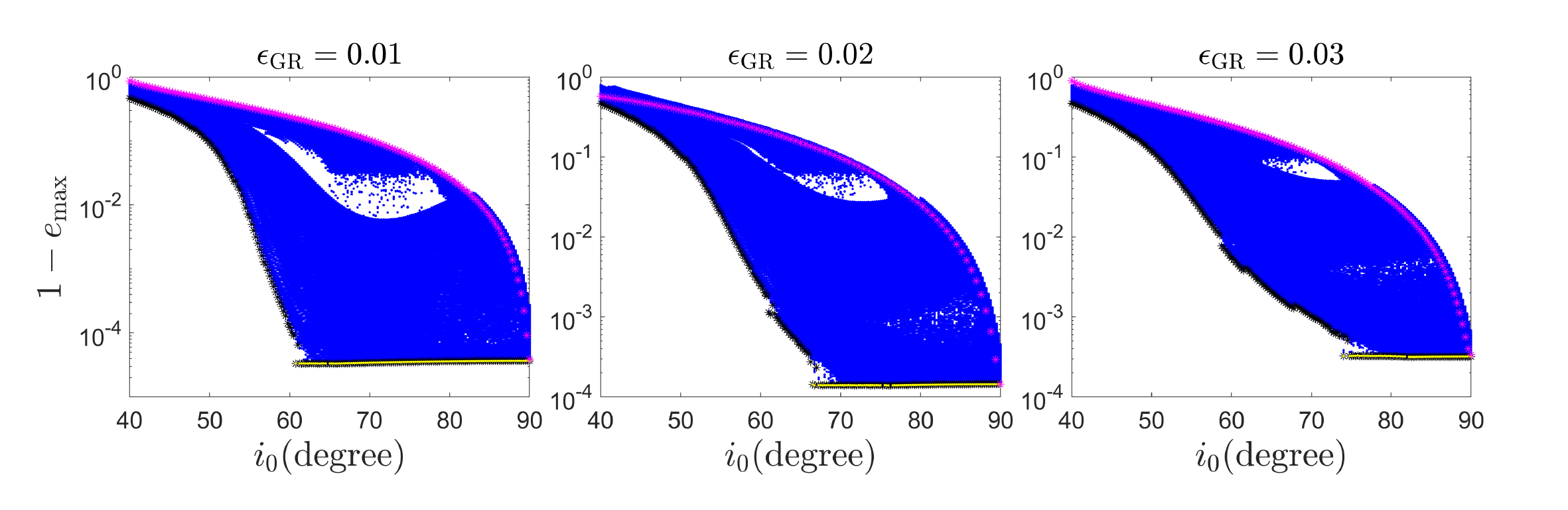}}
\caption{The maximum eccentricity as a function of the initial inclination $i_{0}$. The parameters are the same as in Fig. \ref{figOCT1}. Each blue dot denotes the maximum eccentricity for a set of initial $\omega_{0}$ and $\Omega_{0}$ in the range $[0,2\pi]$. For a given $i_0$, the highest value of the maximum eccentricities obtained from different initial $\omega_0$ and $\Omega_0$ is marked by a black asterisk. The black asterisks are the same as in Fig. \ref{figOCT1}. If the orbit represented by the black asterisk can realize flip, it is additionally marked by a yellow dot. The magenta asterisks are obtained by Eq. (\ref{equa1}).}
\label{figOCT2}
\end{figure*}

The analysis of \citet{liu2015suppression} was restricted to $e_0 \simeq 0$. To examine the ``octupole + GR" model systematically, we numerically integrate the equations of motion over 1500 ZLK cycles (for each parameter set) for different values of $\epsilon_{\rm GR}$, all with $\epsilon_{\rm Oct}=0.02$ and $e_0=0.2$; the results are shown in Fig. \ref{figOCT1}. Note that when $e_0 \neq 0$, $e_{\max}$ depends not only on $i_0$, but may also depend on the initial $\omega_0$ and $\Omega_0$. From Fig. \ref{figOCT1}, we see that the maximum eccentricity is achieved at $\omega_0=0$ and $\Omega_0=0$ when $i_0$ is sufficiently high, but at lower $i_0$, the maximum eccentricity is achieved at other values of $\omega_0$ and $\Omega_0$. Most importantly, we see that for $\epsilon_{\rm GR}=0$, the eccentricity can be excited to extremely high values when $i_0$ is larger than a critical value (see the black asterisks in the top-left panel in Fig. \ref{figOCT1}), and with the increase of $\epsilon_{\rm GR}$, the extreme eccentricity is suppressed to a limiting value. This limiting eccentricity decreases and the critical initial inclination angle increases with increasing $\epsilon_{\rm GR}$.
   
The maximum eccentricity of each orbit with initial angles $\omega_0$ and $\Omega_0$ randomly chosen in the range $[0,2\pi]$ are shown by the blue dots in Fig. \ref{figOCT2}. Other parameters are the same as those in Fig. \ref{figOCT1}. The black asterisks are also the same as the ones in Fig. \ref{figOCT1}, which are the maximums among the blue dots for a given $i_0$. If the orbits represented by the black asterisks can flip, they are additionally marked with yellow dots. For comparison, the maximum eccentricities obtained in the ``quadrupole + GR" model are shown by the magenta asterisks. We see that the maximum eccentricity for orbits which can flip (i.e., the yellow dots) has a limiting value, and is very close to $e_{\rm lim}$. We define the limiting eccentricity for flipping orbits as $e_{\rm flip}$. As $\epsilon_{\rm GR}$ increases, $(1-e_{\rm flip})$ and the critical inclination which allows orbital flips both increase. 

\subsection{Theoretical analysis: ``Theory" of $e_{\rm lim}$}
\begin{figure*}
\centering
{\includegraphics[width=0.99\textwidth]{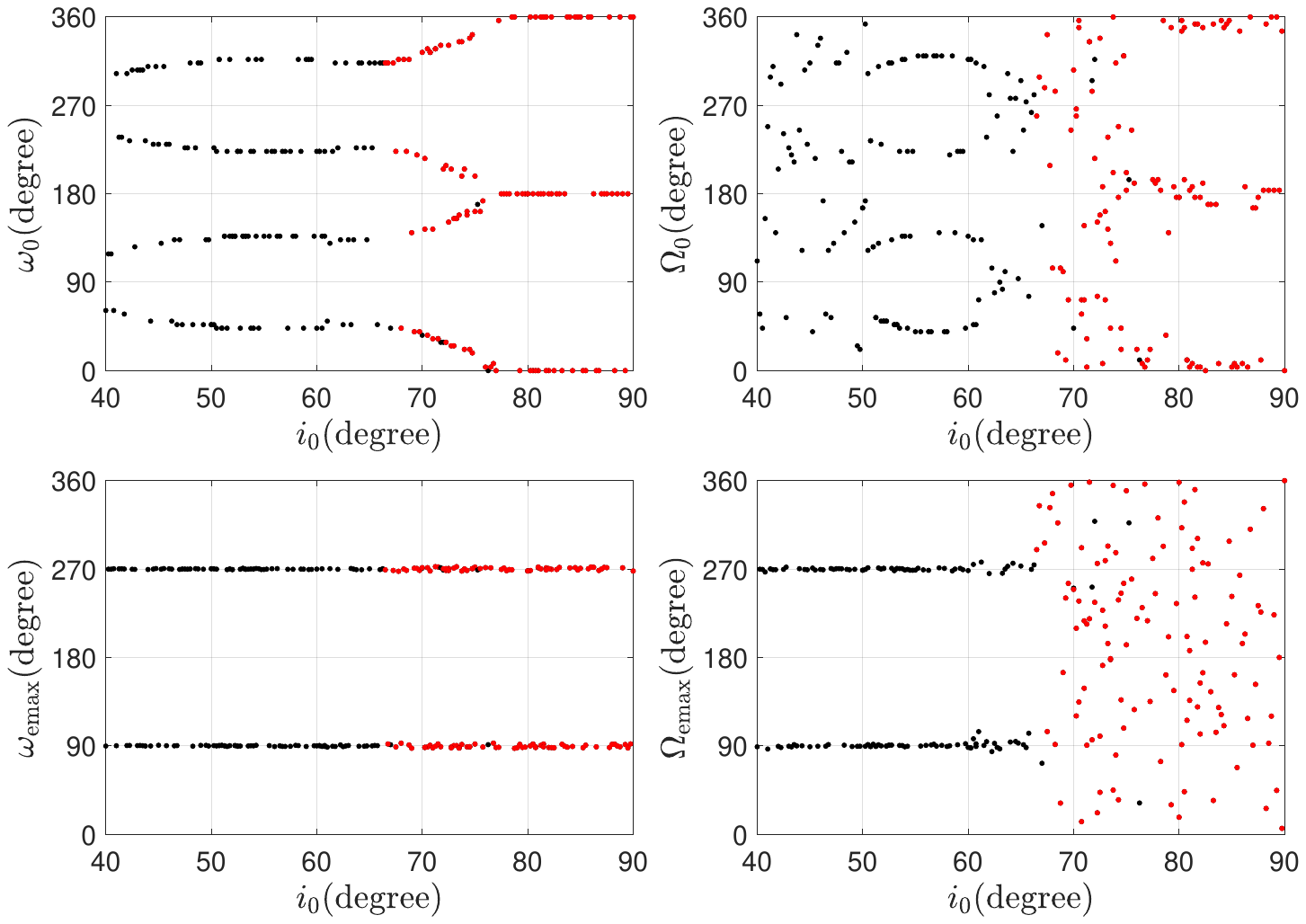}}
\caption{Top: The initial angles corresponding to the black asterisks in the middle panel in Fig. \ref{figOCT2}. Bottom: The angles when the maximum eccentricity is reached. The red dots represent the orbits that can realize flips, and the black dots are orbits that cannot flip.}
\label{figangle}
\end{figure*}

\begin{figure}
\centering
{\includegraphics[width=0.49\textwidth]{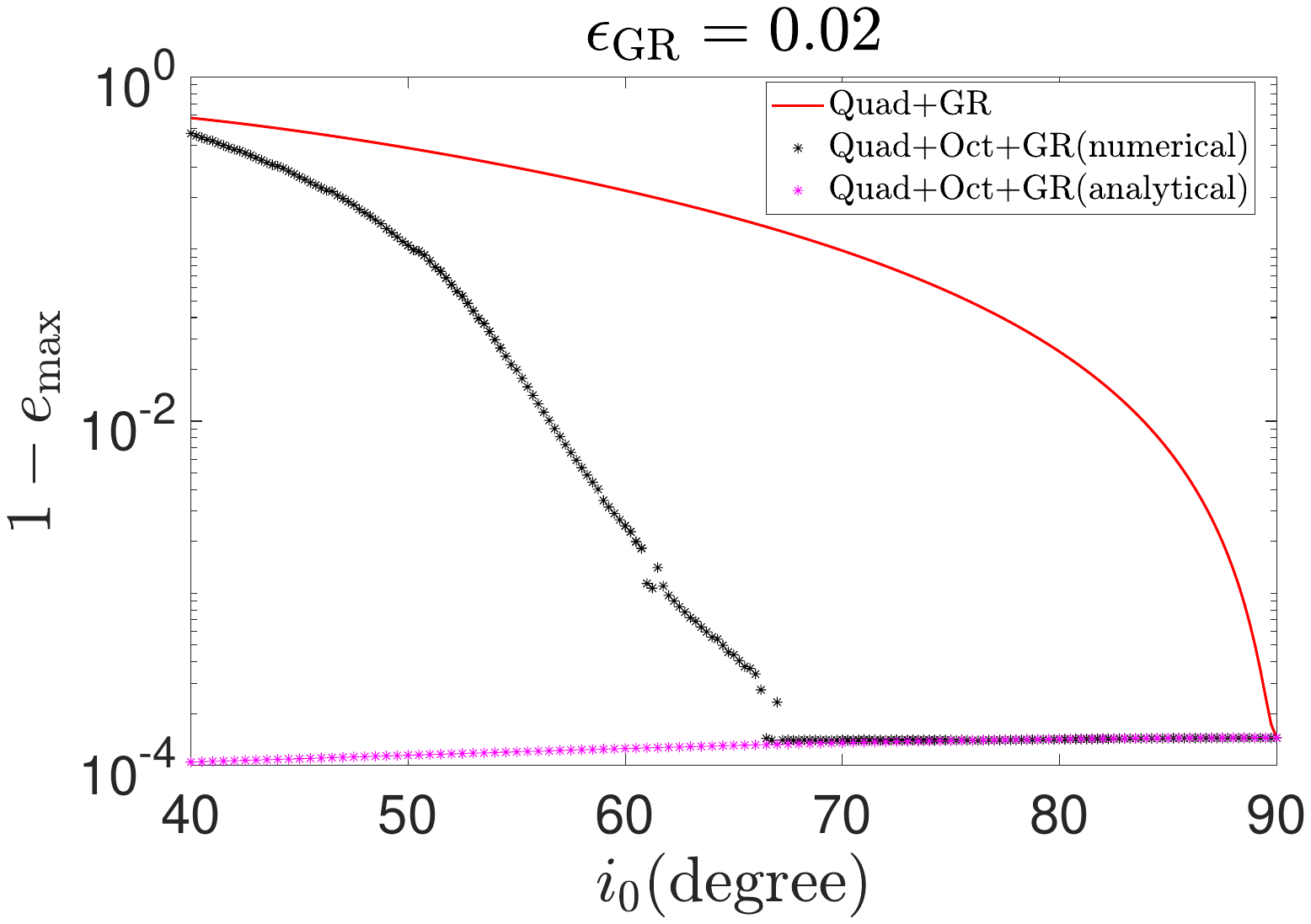}}
\caption{Analytical maximum eccentricity of flipping orbits. The magenta asterisks are $e_{\rm flip}$ obtained from Eq. (\ref{eflip}), the black asterisks are the same as the ones in the middle panel in Fig. \ref{figOCT2}, and the red line are the same as the one in Fig. \ref{fig1}. As can be seen, the analytical results are in good agreement with the numerical ones, and they are consistent with the $e_{\rm lim}$ obtained from Eq. (\ref{elim}).}
\label{figana1}
\end{figure}

\begin{figure}
\centering
{\includegraphics[width=0.49\textwidth]{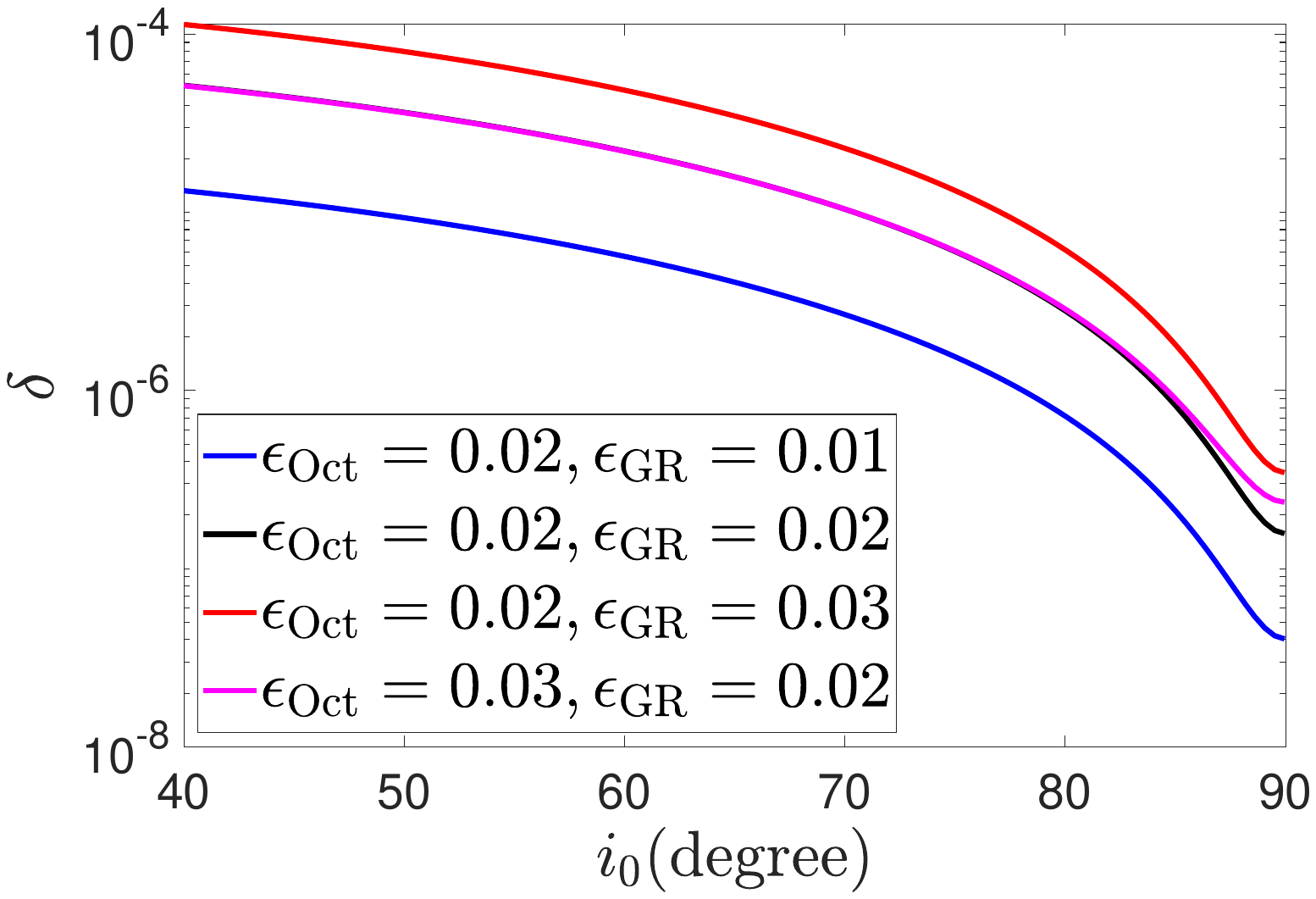}}
\caption{The difference of $e_{\rm flip}$ and $e_{\rm lim}$ as a function of $i_0$ for different $\epsilon_{\rm Oct}$ and $\epsilon_{\rm GR}$.}
\label{figdelta}
\end{figure}

We have shown numerically that the limiting eccentricity (denoted by $e_{\rm flip}$) for flipping orbits in the ``octupole + GR" model is very close to $e_{\rm lim}$. Here we derive $e_{\rm flip}$ analytically and explain why $e_{\rm flip} \simeq e_{\rm lim}$. In the ``octupole + GR" model, the only constant of motion is the total energy $\cal H$, thus
\begin{equation}\label{equa3}
   \begin{array}{l}
      {\cal H}({\omega_{0}},{\Omega_{0}},{e_{0}},{i_{0}})
 = {\cal H}({\omega_{\rm{emax}}},{\Omega_{\rm{emax}}},{e_{\max }},{i_{\rm{emax} }}),
  \end{array}
\end{equation}
where the subscript `0' stands for the initial conditions and the subscript `$\rm{emax}$' stands for the orbital elements when $e=e_{\max}$. At an orbital flip, $i_{\rm emax}=90^{\circ}$. To solve $e_{\rm max}$, it is necessary to know the values of $\omega_0$, $\Omega_0$, $\omega_{\rm emax}$, and $\Omega_{\rm emax}$. We take the black asterisks in the middle panel in Fig. \ref{figOCT2} as an example to study these angles. The top panels in Fig. \ref{figangle} show the initial angels, and the bottom panels show the angles at $e=e_{\max}$. The orbits that can realize flips are denoted by the red dots, and the orbits that cannot flip are denoted by the black dots. When $e=e_{\max}$, the angle $\omega_{\rm emax}$ is always $\pm \pi/2$, and $\Omega_{\rm emax}$ corresponding to orbits without flips is $\pm \pi/2$. For the orbits that can realize flips, $\Omega_{\rm emax}$ is random. This is because $\Omega$ is only contained in the octupole-order term, and when $\omega=\pm \pi/2$ and $i=\pi/2$, the octupole term is equal to 0, regardless of the value of $\Omega$.

Therefore, $e_{\rm flip}$ satisfies the equation
\begin{equation}\label{eflip}
\begin{aligned}
   &{\hat{\cal H}}({e_0},{i_0},{\omega _0},{\Omega _0})\\
 = &{\hat{\cal H}_{{\rm{Quad}}}}({e_{\rm flip}},{i} = \frac{\pi }{2},{\omega} = \pm \frac{\pi }{2})+ {\hat{\Phi} _{\rm GR}}({e_{\rm flip}}).
\end{aligned}
\end{equation}
Furthermore, if we set $\omega_0=\Omega_0=0$ for the flipping orbits (see Fig. \ref{figangle}), we can use Eq. (\ref{eflip}) to obtain $e_{\rm flip}$ as a function of $e_0$ and $i_0$. Fig. \ref{figana1} shows the result for the case of $e_0=0.2$. We see that the analytically determined $e_{\rm flip}$ agrees with the numerical results.

The difference between $e_{\rm lim}$ and $e_{\rm flip}$ is denoted by $\delta$,
\begin{equation}\label{delta}
 \delta  = \frac{{{e_{\rm flip}} - {e_{\rm lim}}}}{{{e_{\rm lim}}}} \ll 1.
\end{equation}
Substituting Eqs. (\ref{elim}) and (\ref{eflip}) into Eq. (\ref{delta}), and expanding the latter for $\delta \ll 1$, we find
\begin{equation}
\begin{aligned}\label{eq_delta}
    \delta  &= \frac{1}{{4e_{\rm lim}^2\left[ {{{9}} - {{{4}{\epsilon _{\rm GR}}}}{{{{\left( {1 - {e_{\rm lim}^2}} \right)}^{-3/2}}}}} \right]}}
   [ - 6  + 6{e_0}^2 \\ 
   & + 16{{\hat{\cal H}}_{\rm Oct}}+{\left( {6 + 9{e_0}^2 - 15{e_0}^2\cos 2{\omega _0}} \right){{\sin }^2}{i_0}} ].
\end{aligned}
\end{equation}
Assuming $\omega_0=0$ and $\Omega_0=0$, Eq. (\ref{eq_delta}) simplifies to
\begin{equation}
 \delta  = \frac{{8{{\hat{\cal H}}_{\rm Oct}} - 3\left( {1 - {e_0}^2} \right){{\cos }^2}{i_0}}}{{2{e_{\rm lim}}^2\left[ {9 - {4{\epsilon _{\rm GR}}}{{{\left( {1 - {e_{\rm lim}}^2} \right)}^{-3/2}}}} \right]}}.
\end{equation}
Fig. \ref{figdelta} shows $\delta$ as a function of $i_0$ for different values of $\epsilon_{\rm Oct}$ and $\epsilon_{\rm GR}$, for the case of $e_0=0.2$. We see that $\delta$ is indeed a small quantity, and becomes smaller with increasing $i_0$. Therefore, for a given initial $e_0$, the limiting (or flipping) eccentricity in the ``octupole + GR" model is well approximated by $e_{\rm lim}$. 

\section{Brown Hamiltonian (BH)}\label{s3}
With lower hierarchy (more precisely, when Eq. \ref{eq_davalid} is violated), the DA model is invalid because the angular momentum of the inner orbit varies significantly in one outer orbital period. The BH can be adopted to account for this variation \citep{luo2016double,grishin2018quasi,tremaine2023hamiltonian}. This is a non-linear perturbation term of the quadrupole-order Hamiltonian, correcting the effects of eccentricity oscillation in one outer orbital period that are neglected in the DA model \citep{tremaine2023hamiltonian,lei2025extensions}.

There are several forms of BH, but they are related by gauge transformations. We adopt the following form of BH in this paper \citep{tremaine2023hamiltonian},
\begin{equation}\label{eq_bh}
\begin{aligned}
{\hat{\cal H}}_{\rm B} = & - \frac{{\rm{3}}}{{{\rm{64}}}}{\epsilon _{\rm B}} \sqrt{1 - e^2} \theta[ 1 + 24e^2 \\
&- (1 - e^2){{\theta }^2} - 15e^2 (1-{\theta }^2){{\sin }^2}\omega  ],
\end{aligned}
\end{equation}
where $\epsilon_{\rm B}$ is a dimensionless parameter measuring the significance of the BH relative to the quadrupole-order term of the DA model,
\[{\epsilon _{\rm B}} =  \frac{P_{\rm out}}{2\pi t_{\rm ZLK}}(3+2e_p^2).\]
The Hamiltonian for the corrected DA model with GR becomes
\begin{equation}
   {\cal H} = \Phi_0({\hat{\cal H}_{\rm Quad}} + {\hat{\cal H}_{\rm Oct}} + {\hat{\cal H}_{\rm B}} + \hat{\Phi}_{\rm GR}).
\end{equation}

\subsection{The quadrupole-order approximation}\label{sbh_q}
\begin{figure}
\centering
{\includegraphics[width=0.49\textwidth]{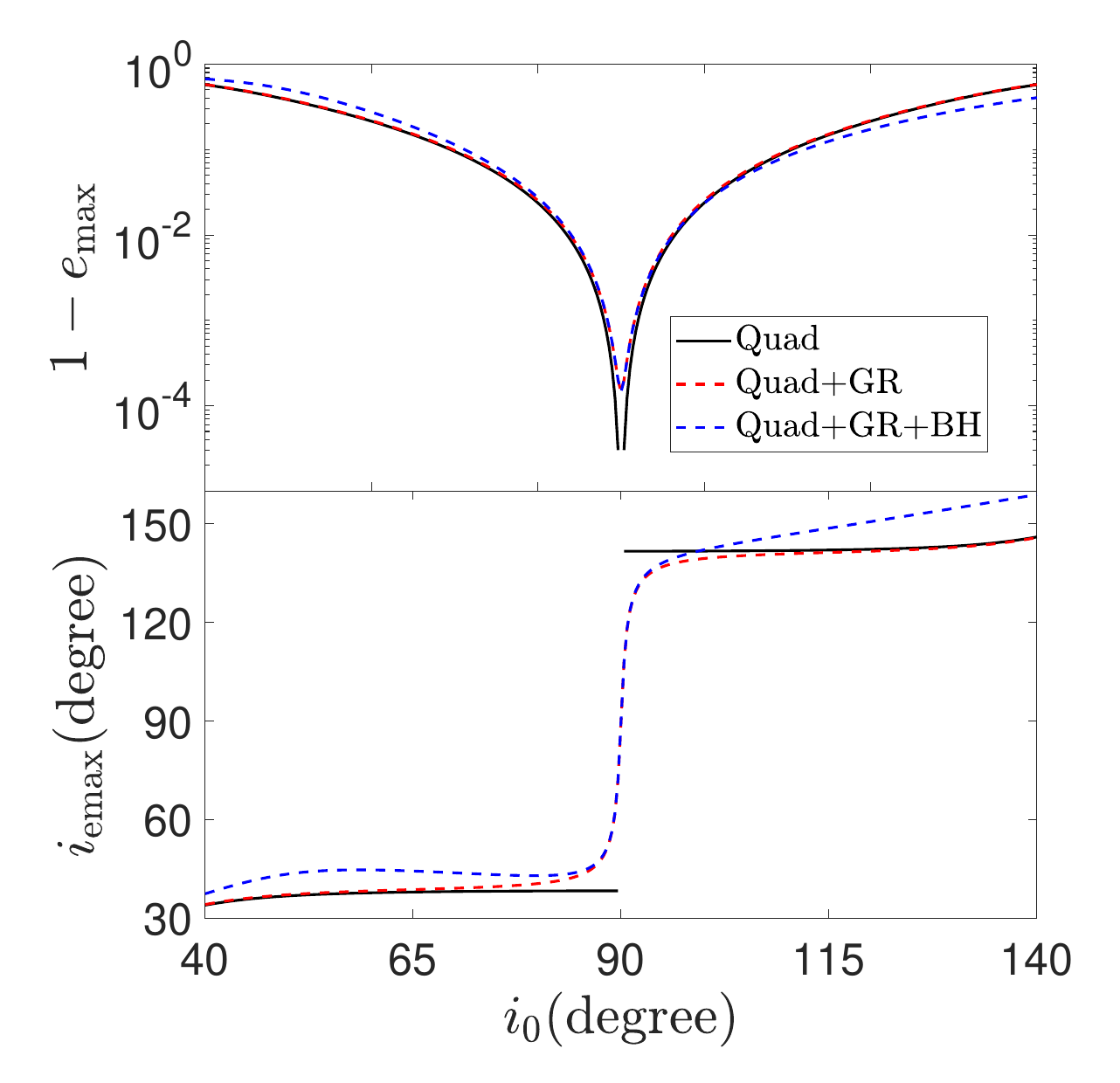}}
\caption{Top panel: The analytical maximum eccentricity as a function of the initial inclination $i_{0}$ for different models, with the initial eccentricity $e_{0}=0.2$. The black solid line is the pure quadrupole model; the red dashed line shows $e_{\max}$ in the ``quadrupole + GR" model with $\epsilon_{\rm {GR}}=0.02$; the blue dashed line shows $e_{\max}$ in the ``quadrupol + GR + BH" model with $\epsilon_{\rm {GR}}=0.02$ and $\epsilon_{\rm {B}}=0.064$. Bottom panel: The corresponding inclination at $e=e_{\max}$.}
\label{figbh_quad}
\end{figure}

The (dimensionless) Hamiltonian model of ``$\hat{\cal{H}}_{\rm Quad}$ + $\hat{\cal{H}}_{\rm B}$ + $\hat{\Phi} _{\rm GR}$" still has one degree of freedom, so the maximum eccentricity in this model can be derived analytically. The total energy and the $z$-component of the inner orbital angular momentum are conserved. According to Eq. (\ref{eq_bh}), the ZLK resonant center is still located at $\omega=\pm \pi/2$, but the dynamics are not symmetric about $i=90^{\circ}$. Given the initial eccentricity and inclination, the maximum eccentricity is obtained when the initial $\omega_0=0$ or $\pi$, and the angle at $e_{\max}$ is $\omega_{\rm emax}=\pm \pi/2$. Therefore, the maximum eccentricity is given by 
\begin{equation}
\begin{array}{l}
    {\hat{\cal H}_{\rm Quad}}({e_{\rm{0}}},{i_{\rm{0}}},{\omega _{\rm{0}}}{\rm{ = 0}}) + {\hat{\Phi} _{\rm GR}}({e_{\rm{0}}}) + {\hat{\cal H}_{\rm B}}({e_{\rm{0}}},{i_{\rm{0}}},{\omega _{\rm{0}}}{\rm{ = 0}})\\
    {\rm{ = }}{\hat{\cal H}_{\rm Quad}}({e_{\max }},{i_{\rm emax}},\omega {\rm{ = }}\pm \frac{\pi }{2}) + {\hat{\Phi} _{\rm GR}}({e_{\max}}) \\
    + {\hat{\cal H}_{\rm B}}({e_{\max }},{i_{\rm emax}},\omega {\rm{ = }}\pm \frac{\pi }{2})\\
\end{array}   
\end{equation}
and
\begin{equation}
    \sqrt {1 - e_0^2} \cos {i_0} = \sqrt {1 - e_{\max }^2} \cos {i_{\rm emax }}.
\end{equation}
Fig. \ref{figbh_quad} shows the analytical maximum eccentricity $e_{\max}$ and the corresponding inclination at $e_{\max}$ as a function of the initial inclination $i_0$ in different one-degree-of-freedom models, for $e_0 =0.2$, $\epsilon_{\rm GR}=0.02$, and $\epsilon_{\rm B}=0.064$. We see that (a) compared to the pure quadrupole model, GR reduces the maximum eccentricity, making ($1-e_{\max}$) non-zero when $i_0$ approaches $90^{\circ}$; (b) compared to the ``quadrupole + GR" model, the BH reduces $e_{\max}$ in the prograde region, and increases $e_{\max}$ in the retrograde region; and (c) BH does not change $e_{\rm lim}$. Point (c) arises because when $i_0=90^{\circ}$, the $z$-component of inner orbital angular momentum equals to 0; thus $\hat{\cal H}_{\rm B}$ vanishes when $i_0=90^{\circ}$, and $e_{\rm lim}$ remains the unchanged.

\subsection{Octupole-order approximation}
We have shown in Section \ref{sbh_q} that BH does not change $e_{\rm lim}$ in the quadrupole approximation. In addition, we have shown in Section \ref{s2} that the ``flipping" eccentricity $e_{\rm flip}$ in the octupole-order approximation with GR can be approximated by  $e_{\rm lim}$. So we expect that BH should not influence $e_{\rm flip}$ in the octupole-order approximation. 
\begin{figure}
\centering
{\includegraphics[width=0.49\textwidth]{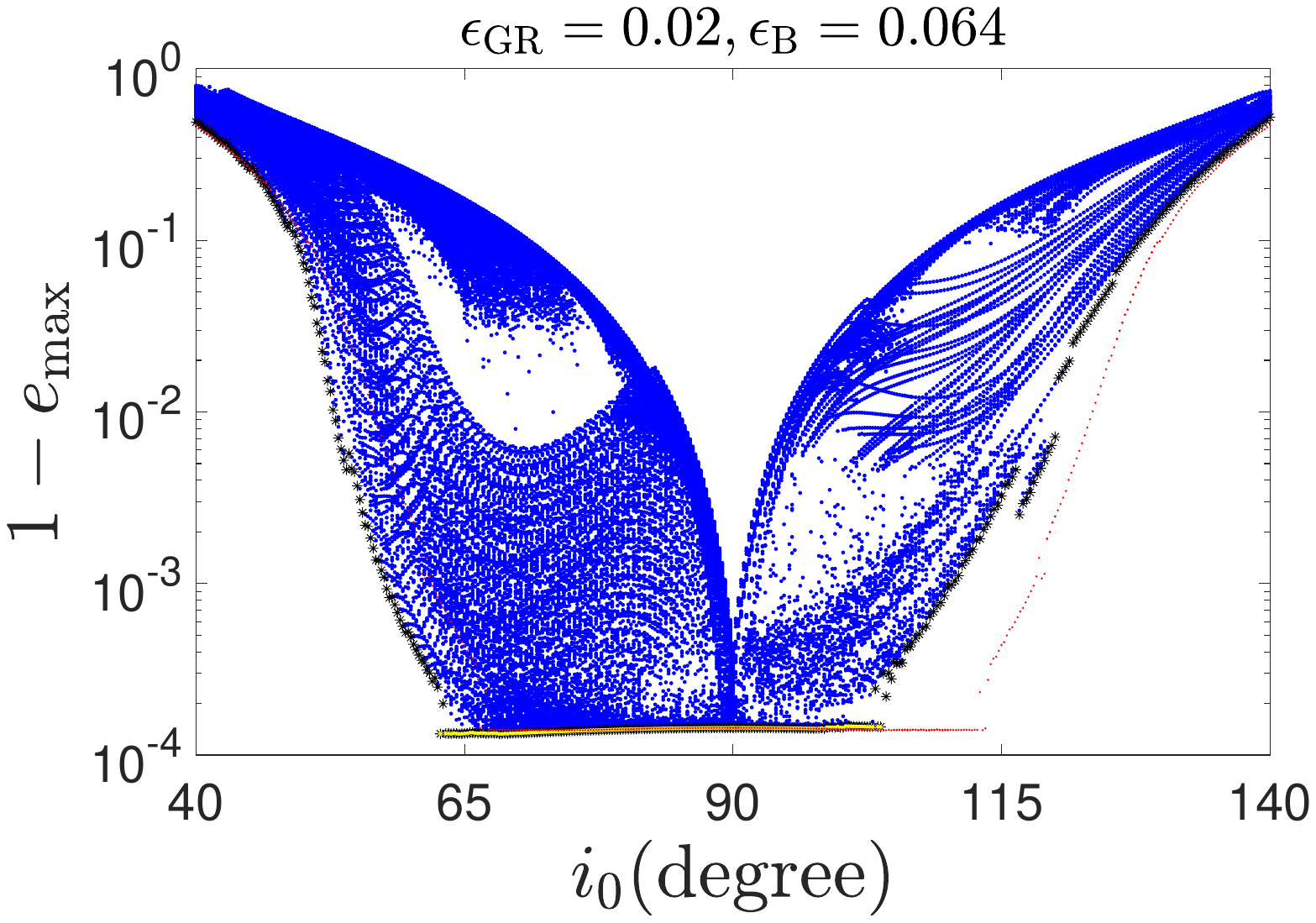}}
\caption{Same as Fig. \ref{figOCT2}, but add BH with $\epsilon_{\rm B}=0.064$. The other parameters are the same as those in the middle panel of Fig. \ref{figOCT2}. The red dots are the same as the black asterisks in the middle panel of Fig. \ref{figOCT2}.}
\label{figbh_oct}
\end{figure}

Fig. \ref{figbh_oct} shows an example of $e_{\max}$ as a function of $i_0$ for the ``$\hat{\cal H}_{\rm Quad}+\hat{\cal H}_{\rm Oct}$ + $\hat{\Phi} _{\rm GR}$ + $\hat{\cal H}_{\rm B}$" model. We set $\epsilon_{\rm B}$ to 0.064, while the other initial parameters are the same as those in the middle panel of Fig. \ref{figOCT2}. We scan the initial angles $\omega_0$ and $\Omega_0$ in the range $[0,2\pi]$, and the corresponding $e_{\max}$ for various $\omega_0$ and $\Omega_0$ values is shown as the blue dot. Among the blue dots, the one with the largest $e_{\max}$ is marked by a black asterisk. In addition, for the black asterisks, the orbits that can realize flip are further marked by yellow dots. We see that including the BH does not change $e_{\rm flip}$.

\section{Single averaging (SA) model}\label{s4}
We have seen in Section \ref{s3} that including the BH does not change the maximum eccentricity of the orbits that can flip. However, \citet{liu2018black} show that the limiting eccentricity in the SA model can even be higher than the one in the DA model (see their Fig. 12). This suggests that the BH model discussed in Section \ref{s3} has its limitations. Here we employ the SA model to study the maximum eccentricity. 

The equations of motion for the SA system are given in \citet{liu2018black}. The inner orbit is specified by the vectors $\boldsymbol{e}(t)$ and $\boldsymbol{J}(t)=\sqrt{1-e^2}\hat{\boldsymbol L}$ (with $\hat{\boldsymbol L}$ the unit vector of the inner orbital angular momentum), and the outer orbit is specified by the position vector $\boldsymbol{r}_{p}(t)$. The evolution equations for $\boldsymbol{e}$ and $\boldsymbol{J}$ are
\begin{equation}
\begin{aligned}
    \frac{{\rm d}{\boldsymbol{J}}}{{\rm d}t}&=\left. \frac{{\rm d}\boldsymbol{J}}{{\rm d}t} \right|_{\text{Quad}}+\left. \frac{{\rm d}\boldsymbol{J}}{{\rm d}t} \right|_{\text{Oct}},\\
    \frac{{\rm d}{\boldsymbol{e}}}{{\rm d}t}&=\left. \frac{{\rm d}\boldsymbol{e}}{{\rm d}t} \right|_{\text{Quad}}+\left. \frac{{\rm d}\boldsymbol{e}}{{\rm d}t} \right|_{\text{Oct}}+\left. \frac{{\rm d}\boldsymbol{e}}{{\rm d}t} \right|_{\text{GR}},
\end{aligned}
\end{equation}
where
\begin{equation}
\begin{aligned}\label{esa_jq}
\left.\frac{{\rm d}\boldsymbol{J}}{{\rm d}t}\right|_{\text{Quad}} = &\frac{3}{2 t_{\mathrm{ZLK}}}\left( \frac{a_p\sqrt{1-e_p^2}}{r_p} \right)^3 \Big[ 5 (\boldsymbol{e} \cdot \hat{\boldsymbol{r}}_{p})\, \boldsymbol{e} \times \hat{\boldsymbol{r}}_{p}\\
 & - (\boldsymbol{J} \cdot \hat{\boldsymbol{r}}_{p})\, \boldsymbol{J} \times \hat{\boldsymbol{r}}_{p} \Big], 
\end{aligned}
\end{equation}
\begin{equation}
\begin{aligned}\label{esa_eq}
\left.\frac{{\rm d}\boldsymbol{e}}{{\rm d}t}\right|_{\text{Quad}} =& \frac{3}{2 t_{\mathrm{ZLK}}}\left( \frac{a_p\sqrt{1-e_p^2}}{r_p} \right)^3  \Big[5 (\boldsymbol{e} \cdot \hat{\boldsymbol{r}}_{p})\, \boldsymbol{J} \times \hat{\boldsymbol{r}}_{p} \\
&- (\boldsymbol{J} \cdot \hat{\boldsymbol{r}}_{p})\, \boldsymbol{e} \times \hat{\boldsymbol{r}}_{p} - 2 \boldsymbol{J} \times \boldsymbol{e} \Big],
\end{aligned}
\end{equation}
and 
\begin{equation}
\begin{aligned}\label{esa_jo}
\left. \frac{{\rm d}\boldsymbol{J}}{{\rm d}t} \right|_{\text{Oct}} &= \frac{15}{16 t_{\mathrm{ZLK}}}\left( \frac{a_p\sqrt{1-e_p^2}}{r_p} \right)^3 \frac{a}{r_p} \Big[ 
10(\boldsymbol{J} \cdot \hat{\boldsymbol{r}}_{p})(\boldsymbol{e} \cdot \hat{\boldsymbol{r}}_{p})\, \boldsymbol{J} \times \hat{\boldsymbol{r}}_{p} \\
&\quad - (1 - 8e^2)\boldsymbol{e} \times \hat{\boldsymbol{r}}_{p} 
+ 5(\boldsymbol{J} \cdot \hat{\boldsymbol{r}}_{p})^2 \boldsymbol{e} \times \hat{\boldsymbol{r}}_{p} \\
&\quad - 35(\boldsymbol{e} \cdot \hat{\boldsymbol{r}}_{p})^2 \boldsymbol{e} \times \hat{\boldsymbol{r}}_{p} \Big], \\
\end{aligned}
\end{equation}
\begin{equation}
\begin{aligned}\label{esa_eo}
\left. \frac{{\rm d}\boldsymbol{e}}{{\rm d}t} \right|_{\text{Oct}} = &\frac{15}{16 t_{\mathrm{ZLK}}}\left( \frac{a_p\sqrt{1-e_p^2}}{r_p} \right)^3 \frac{a}{r_p} \Big[ 
16(\boldsymbol{e} \cdot \hat{\boldsymbol{r}}_{p}) \boldsymbol{J} \times \hat{\boldsymbol{e}} \\
&- (1 - 8e^2)\boldsymbol{J} \times \hat{\boldsymbol{r}}_{p} + 5(\boldsymbol{J} \cdot \hat{\boldsymbol{r}}_{p})^2 \boldsymbol{J} \times \hat{\boldsymbol{r}}_{p} \\
&- 35(\boldsymbol{e} \cdot \hat{\boldsymbol{r}}_{p})^2 \boldsymbol{J} \times \hat{\boldsymbol{r}}_{p} + 10(\boldsymbol{J} \cdot \hat{\boldsymbol{r}}_{p})(\boldsymbol{e} \cdot \hat{\boldsymbol{r}}_{p})\, \boldsymbol{e} \times \hat{\boldsymbol{r}}_{p} \Big].
\end{aligned}
\end{equation}
The motion induced by GR is
\begin{equation}\label{esa_egr}
    \left.\frac{{\rm d}\boldsymbol{e}}{{\rm d}t}\right|_{\text{GR}} = {\dot \omega}_{\rm GR}{\hat {\boldsymbol L}} \times \boldsymbol{e},
\end{equation}
where
\begin{equation}\label{eq_rategr}
{\dot \omega}_{\rm GR}=\frac{\epsilon_{\rm GR}}{t_{\rm  ZLK}}\frac{1}{1-e^2}    
\end{equation}
is the precession rate induced by GR. 
The equation of motion for outer orbit is given by
\begin{equation}
\begin{aligned}
    \mu_{p} \frac{d^2 {\boldsymbol{r}}_{p}}{dt^2}   =& \nabla_{{\boldsymbol{r}}_{p}} \left( \frac{G m_{*} m_p}{r_p} \right) \\
   & - \nabla_{{\boldsymbol{r}}_{p}} \left( \langle \Phi_{\text{Quad}} \rangle + \langle \Phi_{\text{Oct}} \rangle \right),
\end{aligned}
\end{equation}
where $\mu _{p} = m_* m_p/(m_*+m_p)$ is the reduced mass of the outer orbit, and $\langle \Phi_{\text{Quad}} \rangle$ and $\langle \Phi_{\text{Oct}} \rangle$ are the SA potential \citep{liu2018black}.

\begin{figure}
\centering
{\includegraphics[width=0.5\textwidth]{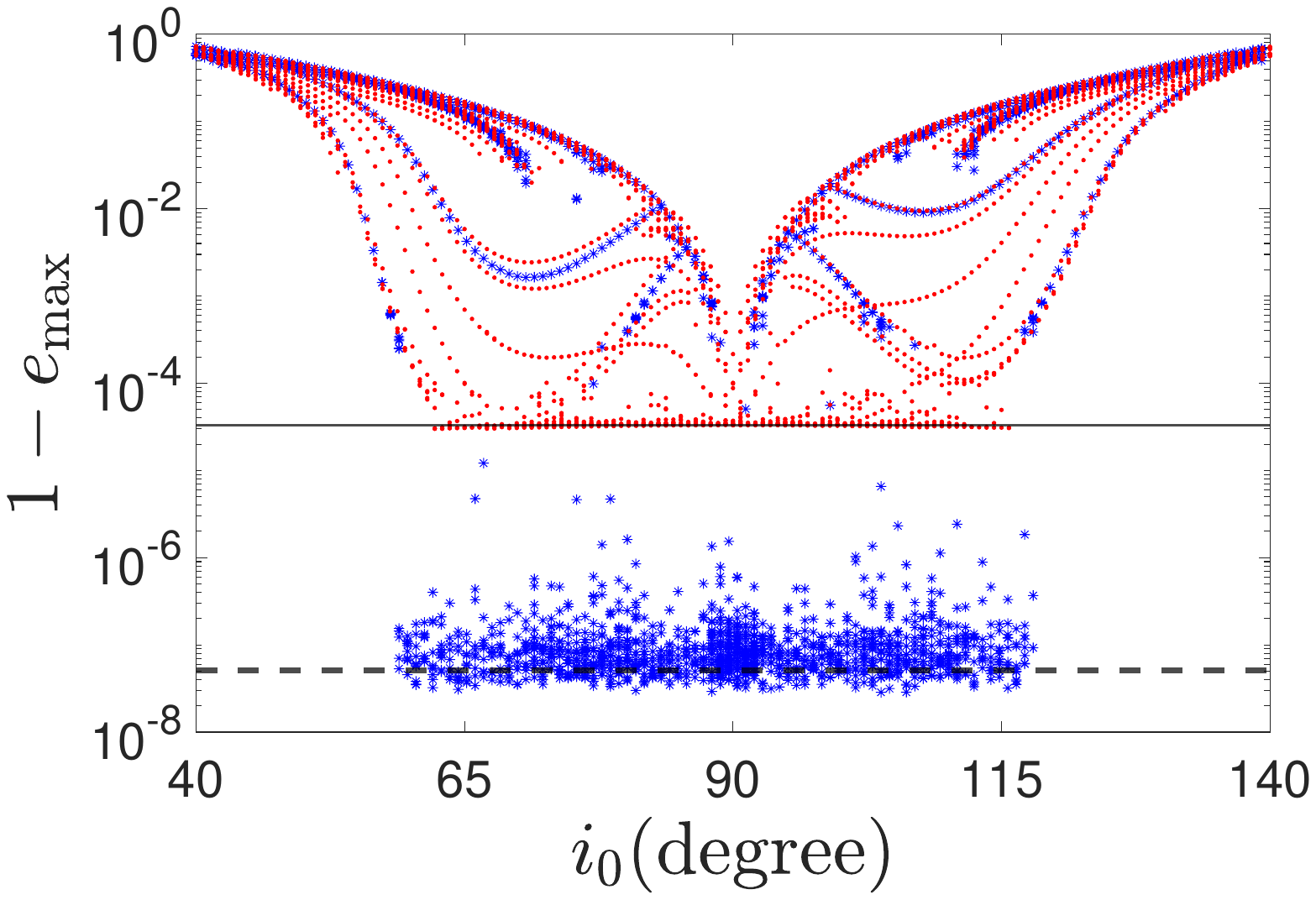}}
\caption{Maximum eccentricity as a function of initial inclination in different models. The parameters are $\epsilon_{\rm Oct}=0.0188$, $\epsilon_{\rm GR}=0.0095$ and $\epsilon_{\rm B}=0.0145$ (with initial inner orbital eccentricity $e_0=0.2$, the outer orbital eccentricity $e_p=0.6$, the semi-major axes of the inner and outer orbit $a=0.2$ au, $a_p=10$ au, the masses $m_*=1M_{\odot}$, $m_p=1M_{\odot}$). The inner orbital parameters $\omega$ and $\Omega$ are scanned in the range $[0,2\pi]$. The red dots represent $e_{\max}$ obtained in the ``DA + BH" model; the blue dots represent $e_{\max}$ obtained in the SA model. The equations of motion are integrated over 1500 ZLK cycles for each parameter set. The solid line is the analytical $e_{\rm lim}$ in the DA model (Eq. \ref{elim}). The dashed line is the analytical estimation of the limiting eccentricity in the SA model (Eq. \ref{eq_elimsa}).}
\label{figSA0}
\end{figure}

\begin{figure}
\centering
{\includegraphics[width=0.49\textwidth]{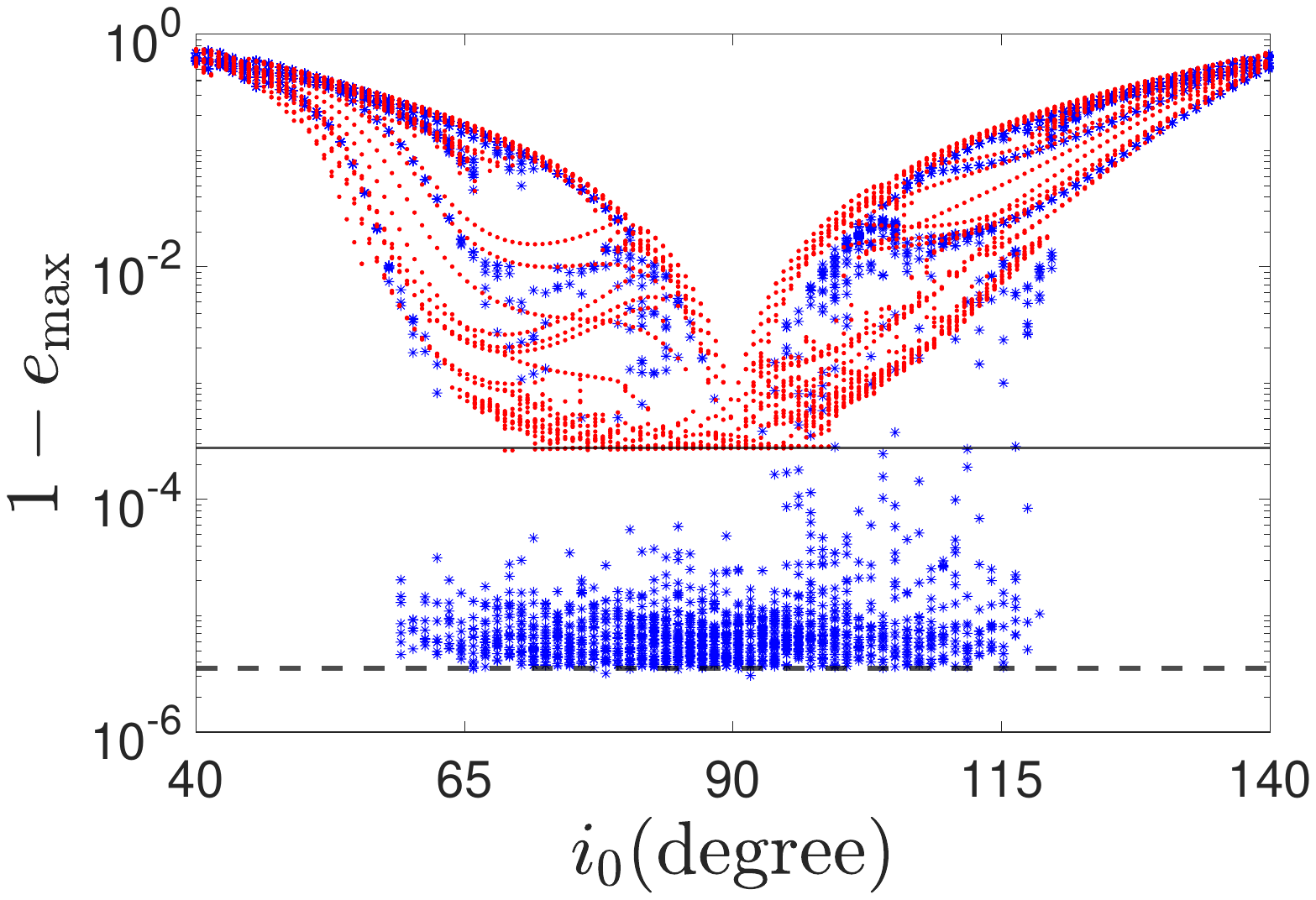}}
\caption{Same as Fig. \ref{figSA0}, but for $\epsilon_{\rm Oct}=0.021$, $\epsilon_{\rm GR}=0.028$ and $\epsilon_{\rm B}=0.073$ (with $e_0=e_p=0.2$, $m_*=m_p=10M_{\odot}$, $a=0.01$ au, $a_p=0.1$au).}
\label{figSA1}
\end{figure}

\begin{figure}
\centering
{\includegraphics[width=0.49\textwidth]{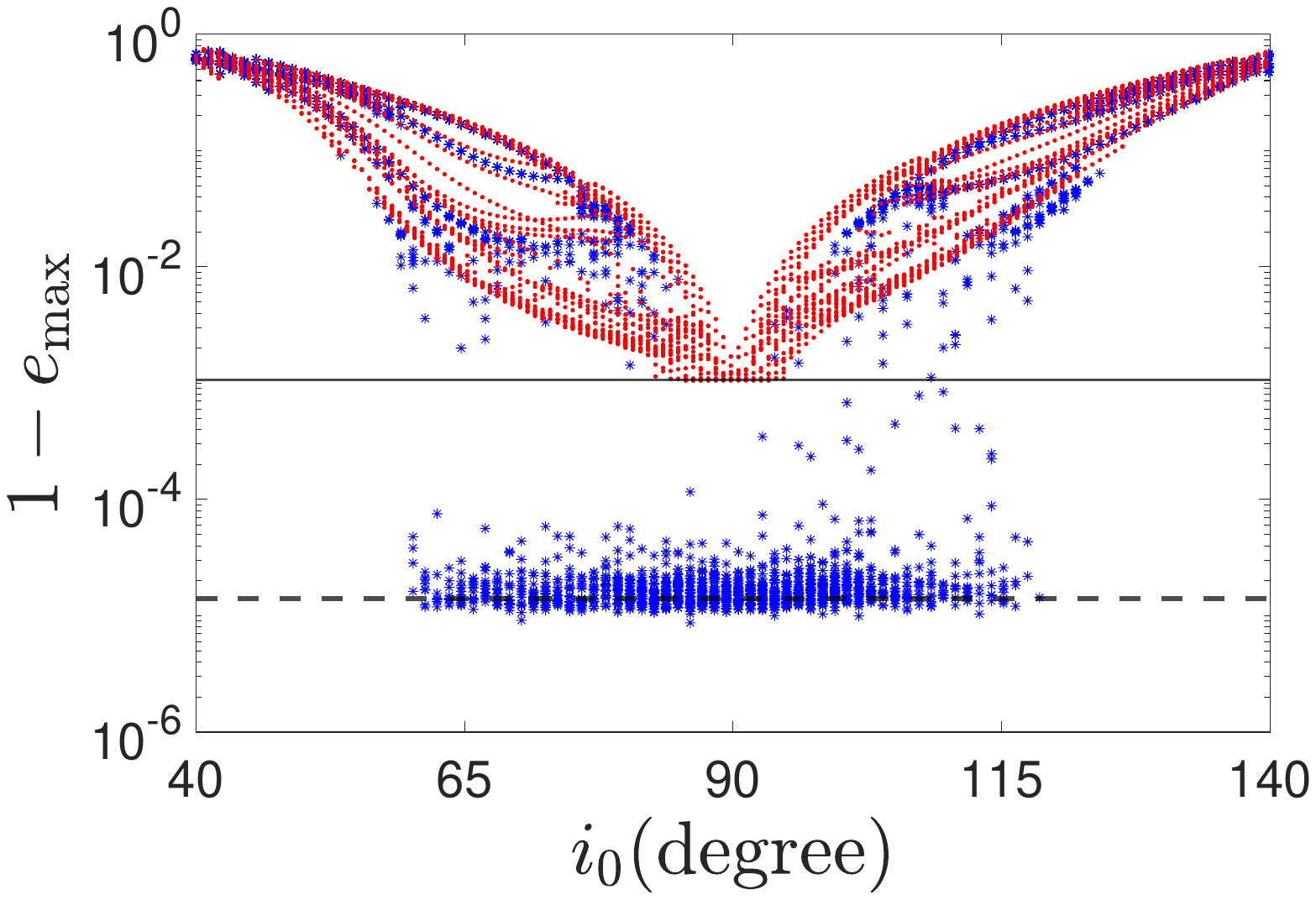}}
\caption{Same as Fig. \ref{figSA0}, but for $\epsilon_{\rm Oct}=0.021$, $\epsilon_{\rm GR}=0.056$ and $\epsilon_{\rm B}=0.073$ (with $e_0=e_p=0.2$, $m_*=m_p=20M_{\odot}$, $a=0.01$ au, $a_p=0.1$au).}
\label{figSA2}
\end{figure}

Figs. \ref{figSA0}, \ref{figSA1} and \ref{figSA2} show examples of the maximum eccentricity as a function of $i_0$ for the ``DA + BH" model and the SA model. The systems chosen in these figures only satisfy the SA approximation but not the DA approximation. We see that given the same initial conditions, the SA model can reach higher eccentricity compared to the DA model. In addition, $e_{\max}$ in the SA model reaches a new limiting eccentricity $e_{\rm lim, SA}$ when $i_0$ is larger than a critical value. 


The new $e_{\rm lim, SA}$ can be understood as follows. Consider the leading (quadrupole) order term in Eq. (\ref{esa_jq}),
\begin{equation}
    \frac{{{\rm d}\boldsymbol J}}{{{\rm d}t}} \simeq \frac{{15}}{{2{t_{\rm ZLK}}}}\left( \frac{a_p\sqrt{1-e_p^2}}{r_p} \right)^3({\boldsymbol e} \cdot \hat{\boldsymbol{r}}_{p})({\boldsymbol e} \times \hat{\boldsymbol{r}}_{p}).
\end{equation}
The maximum rate of change of $\boldsymbol J$ is 
\begin{equation}
\begin{aligned}\label{eq_Jrate}
    \left| {\frac{1}{J}\frac{{{\rm d}\boldsymbol J}}{{{\rm d}t}}} \right|_{\max} &\sim \frac{{15}}{{2{t_{\rm ZLK}} \sqrt {1 - {e_{\max}^2}} }}\left( \frac{a_p\sqrt{1-e_p^2}}{r_p} \right)^3\\
    &\sim \frac{{15{{(1 + {e_p})}^{3/2}}}}{{2{t_{\rm ZLK}}{{(1 - {e_p})}^{3/2}}\sqrt {1 - {e_{\max}}^2} }}.
\end{aligned}
\end{equation}
Equating Eq. (\ref{eq_Jrate}) with ${\dot \omega}_{\rm GR}(e=e_{\max})$, we obtain an estimate for the limiting eccentricity
\begin{equation}\label{eq_elimsa}
(1 - e_{\rm lim,SA}^2)^{1/2} \simeq \frac{2{(1 - {e_p})^{3/2}}}{{15{{(1 + {e_p})}^{3/2}}}}{\epsilon _{\rm GR}}.
\end{equation}
On the other hand, the DA limiting eccentricity is given by Eq. (\ref{elim}) or Eq. (\ref{elim_simplify}). We see that $(1-e_{\rm {lim, SA}})$ is much smaller than $(1-e_{\rm lim})$.
The analytical SA limiting eccentricity is shown as the dashed line in Figs. \ref{figSA0}--\ref{figSA2}, and it agrees well with the numerical result.

\section{Conclusion}\label{s5}
In this work, we have studied the maximum eccentricity attainable by the inner binary under the influence of the perturbation from the tertiary and the short-range effect (taking GR as an example) in the restricted hierarchical three-body system. 

In systems with sufficiently high hierarchy (see Eq. \ref{eq_davalid}), the double averaging (DA) model is a good approximation. Considering the binary-tertiary interaction potential only up to the quadrupole order and including the GR effect, the system has one degree of freedom. For a given $\epsilon_{\rm GR}$ (which measures the strength of the GR effect), the maximum eccentricity $e_{\max}$ as a function of the initial inclination $i_0$ can be obtained analytically (see Eq. \ref{equa1}); for a finite $\epsilon_{\rm GR}$, this $e_{\max}$ approaches a limiting value $e_{\rm lim}$ at $i_0=90^{\circ}$ (see Eq. \ref{elim}). When including the octupole potential, numerical integrations show that the inner orbit can attain a maximum eccentricity nearly equal to $e_{\rm lim}$ for modest initial inclinations ($i_0 < 90^{\circ}$), accompanied by orbital flips. We confirm this numerical finding in the ``octupole + GR" model by analytically calculated $e_{\rm flip}$, the inner orbital eccentricity at which orbital flip occurs, and showing that $e_{\rm flip}$ is indeed very close to $e_{\rm lim}$.

When the system is mildly hierarchical (i.e., Eq. \ref{eq_davalid} is violated), the DA model breaks down. The Brown Hamiltonian (BH) can be introduced as a correction to the DA model, or the single averaging (SA) model (i.e., the binary-tertiary interaction potential is only averaged over the inner orbital period) can be employed. For the DA ``BH + GR" model, we show that $e_{\rm lim}$ and $e_{\rm flip}$ remain unchanged because the BH terms equal to 0 when $i=90^{\circ}$. We use the SA model to numerically study $e_{\rm max}$ in the moderate hierarchical systems, and find that the limiting eccentricity in the SA model is higher than the one in the DA model (see Figs. \ref{figSA0}, \ref{figSA1}, and \ref{figSA2}). This new limiting eccentricity $e_{\rm {lim,SA}}$ in the SA model can be estimated analytically by Eq. (\ref{eq_elimsa}).

Overall, our result shows that while the octupole binary-tertiary potential may induce large eccentricity and orbital flip at modest inclinations, the SRFs always set a limit on the maximum attainable eccentricity, and this limiting eccentricity ($e_{\rm lim}$ or $e_{\rm {lim,SA}}$) is determined by the competition between the quadupole potential and the SRFs. One important issue that is unsolved analytically concerns the critical initial inclination angle $i_{\rm crit}$ above which extreme eccentricity excitation and orbital flip occur. In the DA model, this $i_{\rm crit}$ depends on $\epsilon_{\rm Oct}$ (which measures the strength of the octupole potential; see Eq. \ref{eq_epsilonoct}). Because of the chaotic nature of the orbit at low eccentricities when the octupole potential is significant (large $\epsilon_{\rm Oct}$), it is difficult to determine $i_{\rm crit}$ analytically except when $\epsilon_{\rm Oct} \ll 1$ \citep{katz2011long}. \citet{munoz2016formation} provides a fitting formula for $i_{\rm crit}$ as a function of $\epsilon_{\rm Oct}$ for systems with initial eccentricity $e_0 \simeq 0$. For more general cases ($e_0 \neq$ 0) or when for the SA model, an analytical determination of $i_{\rm cirt}$ remains out of reach.

\begin{acknowledgments}
B.L. acknowledges support from the National Key Research and Development Program of China (No. 2023YFB3002502) and National Natural Science Foundation of China (Grant No. 12433008). X.H. thanks Hanlun Lei and Yubo Su for useful discussions.
\end{acknowledgments}

%\appendix

\bibliography{mybib}{}

@article{vick2019chaotic,
  title={Chaotic tides in migrating gas giants: forming hot and transient warm Jupiters via Lidov--Kozai migration},
  author={Vick, Michelle and Lai, Dong and Anderson, Kassandra R},
  journal={MNRAS},
  volume={484},
  number={4},
  pages={5645--5668},
  year={2019},
  publisher={Oxford University Press}
}

@article{anderson2017eccentricity,
  title={Eccentricity and spin-orbit misalignment in short-period stellar binaries as a signpost of hidden tertiary companions},
  author={Anderson, Kassandra R and Lai, Dong and Storch, Natalia I},
  journal={MNRAS},
  volume={467},
  number={3},
  pages={3066--3082},
  year={2017},
  publisher={Oxford University Press}
}

@article{holman1997chaotic,
  title={Chaotic variations in the eccentricity of the planet orbiting 16 Cygni B},
  author={Holman, Matthew and Touma, Jihad and Tremaine, Scott},
  journal={Nature},
  volume={386},
  number={6622},
  pages={254--256},
  year={1997},
  publisher={Nature Publishing Group UK London}
}

@article{lei2025extensions,
  title={Extensions of Brown Hamiltonian--I. A high-accuracy model for von Zeipel--Lidov--Kozai oscillations},
  author={Lei, Hanlun and Grishin, Evgeni},
  journal={MNRAS},
  volume={540},
  number={3},
  pages={2422--2431},
  year={2025},
  publisher={Oxford University Press}
}

@article{klein2024hierarchical,
  title={Hierarchical three-body problem at high eccentricities= simple pendulum I: octupole},
  author={Klein, Ygal Y and Katz, Boaz},
  journal={Monthly Notices of the Royal Astronomical Society: Letters},
  volume={535},
  number={1},
  pages={L26--L30},
  year={2024},
  publisher={Oxford University Press}
}

@article{grishin2018quasi,
  title={Quasi-secular evolution of mildly hierarchical triple systems: analytics and applications for GW sources and hot Jupiters},
  author={Grishin, Evgeni and Perets, Hagai B and Fragione, Giacomo},
  journal={MNRAS},
  volume={481},
  number={4},
  pages={4907--4923},
  year={2018},
  publisher={Oxford University Press}
}

@article{tremaine2023hamiltonian,
  title={The Hamiltonian for von Zeipel--Lidov--Kozai oscillations},
  author={Tremaine, Scott},
  journal={MNRAS},
  volume={522},
  number={1},
  pages={937--947},
  year={2023},
  publisher={Oxford University Press}
}

@article{liu2018black,
  title={Black hole and neutron star binary mergers in triple systems: Merger fraction and spin--orbit misalignment},
  author={Liu, Bin and Lai, Dong},
  journal={ApJ},
  volume={863},
  number={1},
  pages={68},
  year={2018},
  publisher={IOP Publishing}
}

@article{liu2017spin,
  title={Spin--orbit misalignment of merging black hole binaries with tertiary companions},
  author={Liu, Bin and Lai, Dong},
  journal={ApJL},
  volume={846},
  number={1},
  pages={L11},
  year={2017},
  publisher={IOP Publishing}
}

@article{wu2003planet,
  title={Planet migration and binary companions: The case of HD 80606b},
  author={Wu, Yanqin and Murray, Norm},
  journal={ApJ},
  volume={589},
  number={1},
  pages={605},
  year={2003},
  publisher={IOP Publishing}
}

@article{fabrycky2007shrinking,
  title={Shrinking binary and planetary orbits by Kozai cycles with tidal friction},
  author={Fabrycky, Daniel and Tremaine, Scott},
  journal={ApJ},
  volume={669},
  number={2},
  pages={1298},
  year={2007},
  publisher={IOP Publishing}
}

@article{eggleton2001orbital,
  title={Orbital evolution in binary and triple stars, with an application to SS Lacertae},
  author={Eggleton, Peter P and Kiseleva-Eggleton, Ludmila},
  journal={ApJ},
  volume={562},
  number={2},
  pages={1012},
  year={2001},
  publisher={IOP Publishing}
}

@article{melchor2023tidal,
  title={Tidal Disruption Events from the Combined Effects of Two-body Relaxation and the Eccentric Kozai--Lidov Mechanism},
  author={Melchor, Denyz and Mockler, Brenna and Naoz, Smadar and Rose, Sanaea C and Ramirez-Ruiz, Enrico},
  journal={ApJ},
  volume={960},
  number={1},
  pages={39},
  year={2023},
  publisher={IOP Publishing}
}

@article{munoz2016formation,
  title={The formation efficiency of close-in planets via Lidov--Kozai migration: analytic calculations},
  author={Mu{\~n}oz, Diego J and Lai, Dong and Liu, Bin},
  journal={MNRAS},
  volume={460},
  number={1},
  pages={1086--1093},
  year={2016},
  publisher={Oxford University Press}
}

@article{liu2015suppression,
  title={Suppression of extreme orbital evolution in triple systems with short-range forces},
  author={Liu, Bin and Mu{\~n}oz, Diego J and Lai, Dong},
  journal={MNRAS},
  volume={447},
  number={1},
  pages={747--764},
  year={2015},
  publisher={Oxford University Press}
}

@article{antognini2015timescales,
  title={Timescales of Kozai--Lidov oscillations at quadrupole and octupole order in the test particle limit},
  author={Antognini, Joseph MO},
  journal={MNRAS},
  volume={452},
  number={4},
  pages={3610--3619},
  year={2015},
  publisher={The Royal Astronomical Society}
}

@article{kozai1962secular,
  title={Secular perturbations of asteroids with high inclination and eccentricity},
  author={Kozai, Yoshihide},
  journal={AJ},
  volume={67},
  pages={591--598},
  year={1962}
}

@article{luo2016double,
  title={Double-averaging can fail to characterize the long-term evolution of Lidov--Kozai Cycles and derivation of an analytical correction},
  author={Luo, Liantong and Katz, Boaz and Dong, Subo},
  journal={MNRAS},
  volume={458},
  number={3},
  pages={3060--3074},
  year={2016},
  publisher={Oxford University Press}
}

@article{ford2000secular,
  title={Secular evolution of hierarchical triple star systems},
  author={Ford, Eric B and Kozinsky, Boris and Rasio, Frederic A},
  journal={ApJ},
  volume={535},
  number={1},
  pages={385},
  year={2000},
  publisher={IOP Publishing}
}

@article{lidov1962evolution,
  title={The evolution of orbits of artificial satellites of planets under the action of gravitational perturbations of external bodies},
  author={Lidov, ML},
  journal={P\&SS},
  volume={9},
  number={10},
  pages={719--759},
  year={1962},
  publisher={Elsevier}
}

@article{ito2019lidov,
  title={The Lidov-Kozai Oscillation and Hugo von Zeipel},
  author={Ito, Takashi and Ohtsuka, Katsuhito},
  journal={MEEP},
  volume={7},
  number={1},
  pages={1--113},
  year={2019}
}

@article{naoz2013secular,
  title={Secular dynamics in hierarchical three-body systems},
  author={Naoz, Smadar and Farr, Will M and Lithwick, Yoram and Rasio, Frederic A and Teyssandier, Jean},
  journal={MNRAS},
  volume={431},
  number={3},
  pages={2155--2171},
  year={2013},
  publisher={Oxford University Press}
}

@article{naoz2016eccentric,
  title={The eccentric Kozai-Lidov effect and its applications},
  author={Naoz, Smadar},
  journal={ARA\&A},
  volume={54},
  pages={441--489},
  year={2016},
  publisher={Annual Reviews}
}

@article{lei2022systematic,
  title={A Systematic Study about Orbit Flips of Test Particles Caused by Eccentric Von Zeipel--Lidov--Kozai Effects},
  author={Lei, Hanlun},
  journal={AJ},
  volume={163},
  number={5},
  pages={214},
  year={2022},
  publisher={IOP Publishing}
}

@article{naoz2011hot,
  title={Hot Jupiters from secular planet--planet interactions},
  author={Naoz, Smadar and Farr, Will M and Lithwick, Yoram and Rasio, Frederic A and Teyssandier, Jean},
  journal={Natur},
  volume={473},
  number={7346},
  pages={187--189},
  year={2011},
  publisher={Nature Publishing Group}
}

@article{anderson2016formation,
  title={Formation and stellar spin-orbit misalignment of hot Jupiters from Lidov--Kozai oscillations in stellar binaries},
  author={Anderson, Kassandra R and Storch, Natalia I and Lai, Dong},
  journal={MNRAS},
  volume={456},
  number={4},
  pages={3671--3701},
  year={2016},
  publisher={Oxford University Press}
}

@article{petrovich2015steady,
  title={STEADY-STATE PLANET MIGRATION BY THE KOZAI--LIDOV MECHANISM IN STELLAR BINARIES},
  author={Petrovich, Cristobal},
  journal={ApJ},
  volume={799},
  number={1},
  pages={27},
  year={2015},
  publisher={IOP Publishing}
}

@article{dawson2018origins,
  title={Origins of hot Jupiters},
  author={Dawson, Rebekah I and Johnson, John Asher},
  journal={ARA\&A},
  volume={56},
  pages={175--221},
  year={2018},
  publisher={Annual Reviews}
}

@article{lithwick2011eccentric,
  title={The eccentric Kozai mechanism for a test particle},
  author={Lithwick, Yoram and Naoz, Smadar},
  journal={ApJ},
  volume={742},
  number={2},
  pages={94},
  year={2011},
  publisher={IOP Publishing}
}

@article{katz2011long,
  title={Long-term cycling of Kozai-Lidov cycles: extreme eccentricities and inclinations excited by a distant eccentric perturber},
  author={Katz, Boaz and Dong, Subo and Malhotra, Renu},
  journal={PhRvL},
  volume={107},
  number={18},
  pages={181101},
  year={2011},
  publisher={APS}
}

@article{von1910application,
  title={Sur l'application des s{\'e}ries de M. Lindstedt {\`a} l'{\'e}tude du mouvement des com{\`e}tes p{\'e}riodiques},
  author={von Zeipel, Hugo},
  journal={Astron. Nachr.},
  volume={183},
  pages={345},
  year={1910}
}

@article{krymolowski1999studies,
  title={Studies of multiple stellar systems—II. Second-order averaged Hamiltonian to follow long-term orbital modulations of hierarchical triple systems},
  author={Krymolowski, Y and Mazeh, T},
  journal={MNRAS},
  volume={304},
  number={4},
  pages={720--732},
  year={1999},
  publisher={Blackwell Science Ltd Oxford, UK}
}

@article{antonini2012secular,
  title={Secular evolution of compact binaries near massive black holes: gravitational wave sources and other exotica},
  author={Antonini, Fabio and Perets, Hagai B},
  journal={ApJ},
  volume={757},
  number={1},
  pages={27},
  year={2012},
  publisher={IOP Publishing}
}
\bibliographystyle{aasjournalv7}

\end{document}